\documentclass[pdflatex,sn-mathphys-num]{sn-jnl}

\usepackage{graphicx}%
\usepackage{multirow}%
\usepackage{amsmath,amssymb,amsfonts}%
\usepackage{amsthm}%
\usepackage{mathrsfs}%
\usepackage[title]{appendix}%
\usepackage{xcolor}%
\usepackage{textcomp}%
\usepackage{manyfoot}%
\usepackage{booktabs}%
\usepackage{algorithm}%
\usepackage{algorithmicx}%
\usepackage{algpseudocode}%
\usepackage{listings}%
\usepackage{float}
\usepackage{siunitx}

\theoremstyle{thmstyleone}%
\theoremstyle{thmstyletwo}%

\theoremstyle{thmstylethree}%

\begin{document}

\title{First Direct Observation of a Wakefield Generated with Structured Light}


\author[1,3,*]{Aaron~Liberman}
\author[1,3]{Anton~Golovanov}
\author[1]{Slava~Smartsev}
\author[1]{Sheroy~Tata}
\author[2]{Igor~A.~Andriyash}
\author[1]{Salome~Benracassa}
\author[1]{Eitan~Y.~Levine}
\author[1]{Eyal~Kroupp}
\author[1]{Victor~Malka}

\affil[1]{Department of Physics of Complex Systems, Weizmann Institute of Science, Rehovot 7610001, Israel}
\affil[2]{Laboratoire d’Optique Appliquée, ENSTA Paris, CNRS, Ecole polytechnique, Institut Polytechnique de Paris, 828 Bd des Maréchaux, 91762 Palaiseau, France}
\affil[3]{These authors contributed equally to this work}

\affil[*]{Corresponding author: aaronrafael.liberman@weizmann.ac.il}

\abstract{\textbf{The use of structured light to control the phase velocity of the wake in laser-wakefield accelerators has generated significant interest for its ability to mitigate electron dephasing. Combining the diffraction-free properties of Bessel beams with spatio-temporal shaping of the pulse promises to enable acceleration with an unprecedented combination of long acceleration lengths and high gradients. This would facilitate the acceleration of electrons to energies above 100\,GeV in existing laser facilities. In-depth understanding of the physical mechanisms involved is critical to achieving dephasing-free electron acceleration. Here we present the first experimental observation of wakefields generated by beams that were spatio-temporally sculpted and then focused with a long-focal-depth mirror, known as an axiparabola, which generates a quasi-Bessel beam. The resulting wakefield was imaged using femtosecond relativistic electron microscopy. Novel insights into this minimally explored regime include mapping the wakefield development over the focal depth and studying the effects of spatio-temporal manipulations of the beam on the structure and phase velocity of the wakefield. Such insights pave the way towards realizing the potential of structured-light based solutions to dephasing in laser-wakefield acceleration.} }

\keywords{Laser-Wakefield Acceleration, Phase-Locked Laser-Wakefield Acceleration, Luminal Acceleration, Structured Light, Flying Focus, Axiparabola, Dephasingless Laser-Wakefield Acceleration, Spatio-Temporal Couplings, Femtosecond Relativistic Electron Microscopy, Particle-in-Cell Simulations}

\maketitle

Since the pioneering idea by T. Tajima and J.M. Dawson in 1979 \cite{Tajima_PRL_1979}, laser-wakefield accelerators (LWFAs) have demonstrated their ability to produce high quality, mono-energetic electron beams in a fraction of the length required by standard RF accelerators  \cite{Mangles_Nature_2004,Geddes_Nature_2004,Faure_Nature_2004}. LWFAs show significant promise in applications varying from novel cancer therapy treatments \cite{Glinec_Med_Phys_2006}, to non-destructive material testing \cite{Glinec_PRL_2005}, and to the generation of free-electron lasers \cite{Wang_Nature_2021,Labat_NaturePhotonics_2022}. There exist, however, a few challenges which limit the achievement of ever higher electron energies and more efficient accelerators \cite{Esarey_ReviewOfModernPhysics_2009}. Among these is the dephasing limit---in which the electrons trapped inside the wakefield outpace the wakefield and cease to be accelerated---and the diffraction limit, in which the laser diffracts and is no longer intense enough to drive a wakefield \cite{Joshi_Nature_1984,Esarey_ReviewOfModernPhysics_2009}. 

Several solutions have been proposed for overcoming the dephasing limit, including rephasing the electrons with a density ramp \cite{Sprangle_PRE_2001,Guillaume_PRL_2015}, multi-staged LWFAs \cite{Leemans_PhysicsToday_2009}, and lowering the plasma density in order to increase the phase velocity of the laser in plasma \cite{Leemans_NaturePhysics_2006,Gonsalves_PRL_2019}. Each of these, however, comes with challenges. Rephasing, while yielding a boost in electron energy \cite{Guillaume_PRL_2015} is limited in the amount of extra acceleration it facilitates. Multi-staged LWFAs are experimentally very complex and timing them correctly is demanding. Lowering the plasma density in turn lowers the acceleration gradient, requiring longer acceleration lengths and, thus, the guiding of the laser pulse over multiple Rayleigh lengths \cite{Sprangle_1988_APL_53_2146,Leemans_NaturePhysics_2006}.

A promising way to overcome the dephasing limit is by modifying the velocity with which the intensity peak of the laser driver, and thus the wakefield, propagates along the optical axis \cite{Debus_PRX_2019}. The use of structured light to manipulate this velocity has generated significant interest within the scientific community \cite{Sainte-Marie_Optica_2017,Froula_NaturePhotonics_2018,Caizergues_NaturePhotonics_2020,Palastro_PRL_2020,Ambat_OpticsExpress_2023,Miller_ScientificReports_2023,Liberman_OL_2024,Pigeon_OE_2024}. Theoretical solutions such as colliding two tilted laser pulses \cite{Debus_PRX_2019} or combining longitudinal chromatism with controlled spectral phase \cite{Sainte-Marie_Optica_2017,Froula_NaturePhotonics_2018} have been explored. Perhaps the most experimentally simple and promising method for achieving dephasingless acceleration relies on a combination of spatio-temporally shaping the beam in the near-field and focusing the laser-pulse with a custom optic that generates a quasi-Bessel beam \cite{Caizergues_NaturePhotonics_2020,Palastro_PRL_2020}.

The most suitable optical element that gives the necessary focusing properties while maintaining compatibility with high-power ultra-short pulses is the axiparabola, a long-focal-depth mirror which combines a parabolic focusing term with a controlled spherical aberration \cite{Smartsev_OpticsLetters_2019, Oubrerie_JoO_2022}. The combination of spatio-temporal pulse sculpting and focusing with the axiparabola promises to overcome both beam diffraction and electron dephasing \cite{Caizergues_NaturePhotonics_2020,Palastro_PRL_2020}. Beam diffraction is tackled by utilizing the extended focal depth of the axiparabola and the diminished diffraction seen in Bessel beams \cite{Smartsev_OpticsLetters_2019}. Dephasing, meanwhile, is addressed by combining the axiparabola's inherent modification of the velocity of the intensity peak with a manipulation of the spatio-temporal couplings of the incident beam \cite{Caizergues_NaturePhotonics_2020,Palastro_PRL_2020,Liberman_OL_2024,Ambat_OpticsExpress_2023,Pigeon_OE_2024}. This combination enables the tuning of the velocity of the wakefield, allowing it to be phase-locked to the trapped electrons \cite{Caizergues_NaturePhotonics_2020,Palastro_PRL_2020}.

Much research effort has been dedicated to studying the axiparabola-based approach in the hope of unlocking this potential \cite{Smartsev_OpticsLetters_2019,Caizergues_NaturePhotonics_2020,Palastro_PRL_2020,Oubrerie_JoO_2022,Geng_PoP_2022,Ambat_OpticsExpress_2023,Miller_ScientificReports_2023,Ramsey_PRA_2023,Geng_ChinesePhys_2023,Liberman_OL_2024,Pigeon_OE_2024}. Simulation results show the ability to achieve electron energies in excess of 100\,GeV in existing high power laser facilities \cite{Caizergues_NaturePhotonics_2020}. However, experimental results have, so far, lagged behind \cite{Liberman_CLEO_2024}. 

A significant barrier is the novelty of the wakefield generated by these structured pulses, which differs significantly from the standard, parabola-focused wakefield. These LWFAs are not, as in the standard case, formed by a whole beam focused to a single point but rather by the focusing of different annular sections of the beam to different points along the optical axis. Thus, much of the conventional intuition of the wakefield no longer applies and the physics involved is significantly altered. Given the highly nonlinear nature of the laser--plasma interaction, which involves field ionization and the formation of a nonlinear wakefield, only a detailed look at the wakefield evolution inside of the plasma can yield an accurate picture of the mechanisms involved. 

This paper introduces the first direct observations of the axiparabola-generated wakefield, taken using the Femtosecond Relativistic Electron Microscopy (FREM) technique \cite{Zhang_ScientificReports_2016,Wan_ScienceAdvances_2024,Levine_PRR_2025}. In FREM, a short, bright, relativistic electron bunch, generated by a second LWFA, provides a probe capable of resolving micron-scale, femtosecond-duration structures inside of the plasma wave \cite{Wan_NaturePhysics_2022}. Taking advantage of the high spatial and temporal resolution, FREM was used to provide detailed pictures of the axiparabola-generated wakefield. The images were taken at different points along the evolution of the wakefield and with different spatio-temporal couplings. The results were confirmed by particle-in-cell (PIC) simulations which recreate both the axiparabola-generated wakefield and the electron probe images of this wakefield. The PIC simulations also enhance the understanding of the experimental results, allowing for the extraction, \textit{in situ}, of the plasma wavelength and density. The excited wakefield is shown to simultaneously manifest properties of both non-linear and linear wakefields in a single shot. 
Modifying the pulse-front curvature (PFC)---a spatio-temporal coupling that gives a radially dependent pulse delay which depends quadratically on the radius---is used to study the influence of changing the spatio-temporal pulse-front, and thus the wakefield velocity, on the structure of the wakefield.
The combination of direct images and simulations yields critical insight into the structure and evolution of this novel wakefield, and helps shed light onto how such wakefields could be used to make dephasingless acceleration a reality.

\section*{Results}\label{Results}
\subsection*{Experimental Setup}

 \begin{figure*}[t!]
		\centering
		\includegraphics[width=\linewidth]{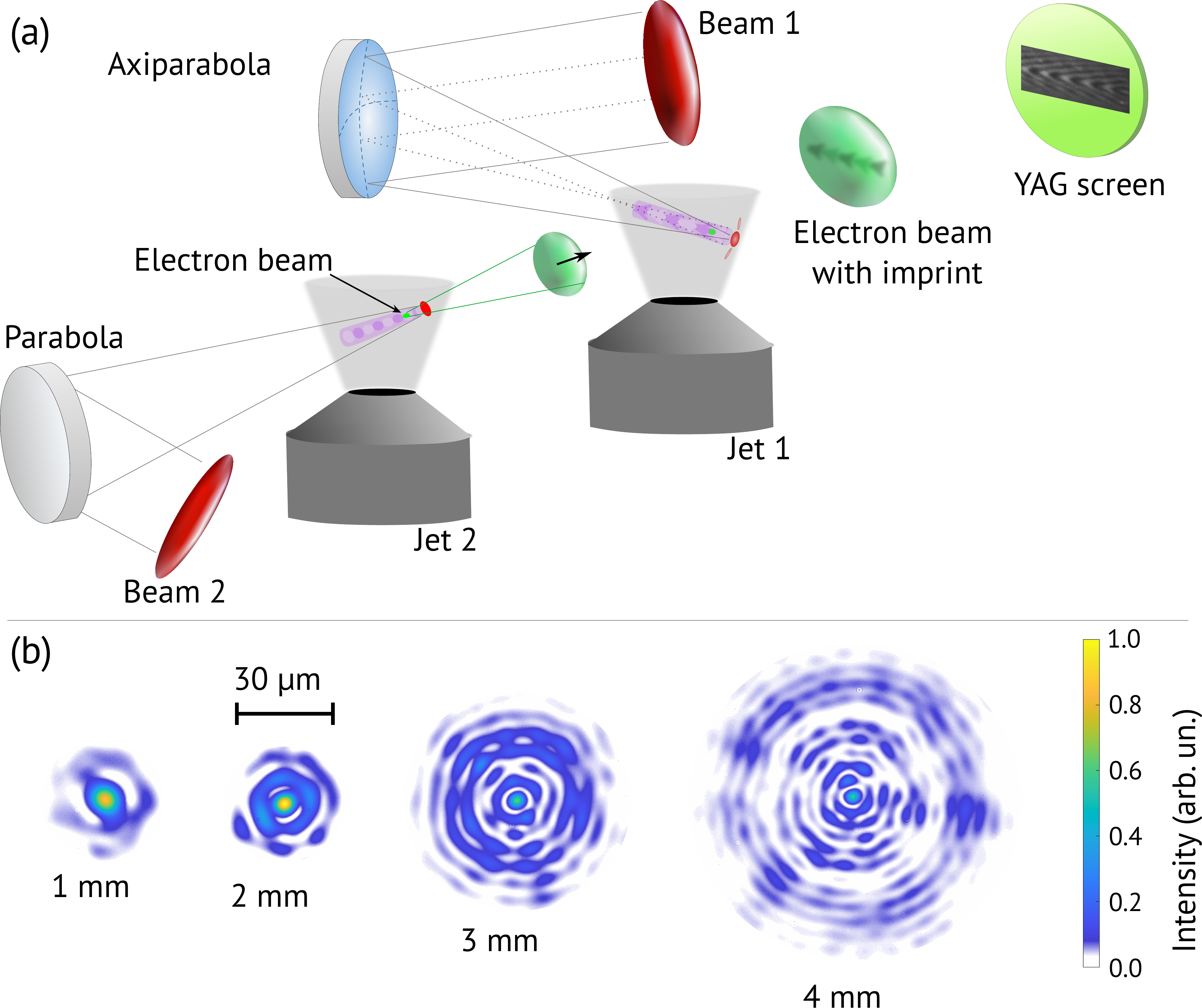}
		\caption{\label{fig:Setup} \small (a) Schematic of the FREM experimental setup. Beam 1 is focused by the axiparabola onto jet 1, generating the novel laser wakefield. Beam 2 is focused by a parabola onto jet 2, generating a second laser wakefield which accelerates a femtosecond duration electron bunch, known as the ``probe electron bunch''. This probe bunch is then allowed to expand and impinges onto the axiparabola-generated wakefield, which imposes momentum changes onto the probe electrons. After further propagation, these momentum changes turn into density modulations which are then imaged when the electrons hit a YAG screen. (b) 2D axiparabola focal spots, measured in vacuum, at specified points along the focal depth relative to the beginning of the focal line \SI{480}{mm} away from the axiparabola.}
\end{figure*}

The experiment was conducted using the HIGGINS $2 \times \SI{100}{TW}$ laser system at the Weizmann Institute of Science \cite{Kroupp_MRE_2022}. The laser system provides two temporally synchronized 2.5\,J, 27\,fs laser pulses. During this experiment, each beam delivered  1\,J on target. Prior to compression, beam 1 was passed through a special refractive doublet designed to modify the pulse-front curvature (PFC) of the beam, thus spatio-temporally shaping the beam in the near-field. Details of the pulse-front curvature measurement and the system used to manipulate the PFC can be found in Ref.~\cite{Smartsev_JoO_2022,Liberman_OL_2024}. 

Figure \ref{fig:Setup} (a) shows a schematic of the experimental setup used to obtain the FREM images. In the experiment, beam 1 was focused by a 480\,mm nominal focal length, 5\,mm focal depth, $10^{\circ}$ off-axis axiparabola onto a supersonic 15\,mm-long slit nozzle, shown in figure \ref{fig:Setup} as ``Jet 1''. The focused beam then generated the structured-light wakefield. Beam 2, meanwhile, was focused by a 1.5\,m-focal-length off-axis parabolic mirror onto a supersonic converging-diverging nozzle, shown as ``Jet 2''. Beam 2 generated a wakefield in Jet 2 and accelerated an electron bunch with femtosecond duration, referred to as the ``probe electron bunch''. A sample spectrum and charge of the probe bunch can be seen in the Lanex scintillator image in extended data figure \ref{fig:B2_Lanex}. This bunch was then allowed to propagate for 10\,cm in order to spatially expand the bunch before it impinged onto the axiparabola-generated wakefield at near normal incidence. The electro-magnetic field inside of the wake gave momentum kicks to the probe electrons. After a further 7\,mm of propagation in vacuum, these momentum changes turned into electron density modulations. The probe beam, imprinted with the density modulations, then hit a YAG screen and the subsequent radiation was imaged. 

Figure \ref{fig:Setup} (b) shows the 2D focal spots of the axiparabola-focused beam in vacuum at different focal depth values. The evolution of the Bessel ring structures as well as the preservation of a small central focal spot, which becomes progressively smaller as the focal depth increases, can be clearly seen. More details of the particular axiparabola that was used can be found in Ref.~\cite{Liberman_OL_2024}. 

\subsection*{The Structured-Light Wakefield}

 \begin{figure*}[t!]
		\centering
		\includegraphics[width=\linewidth]{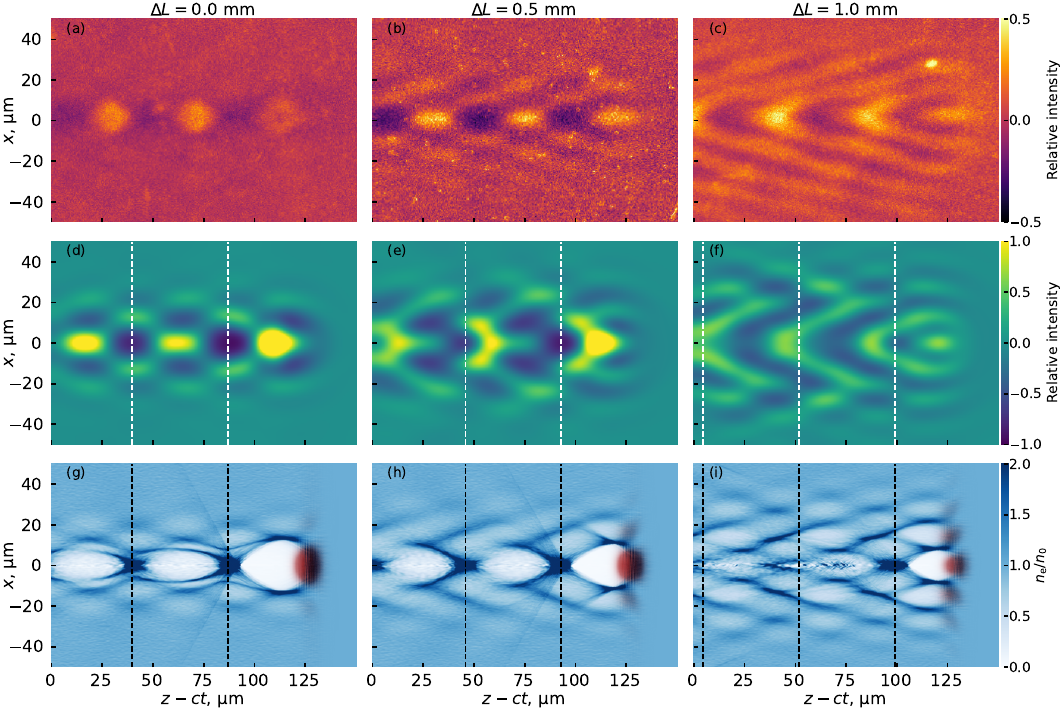}
		\caption{\label{fig:Data} \small (a--c) Experimentally obtained FREM images at different relative points $\Delta L$ along the focal depth showing the development of the V-shaped structure. (d--f) Simulated FREM images for similar experimental conditions and at similar points in the nozzle. Image (d) corresponds to the depth of 3.95\,mm inside the nozzle (0.95\,mm after the beginning of the focal line) in the PIC simulations. (g--i) Relative electron density distribution $n_\mathrm{e} / n_0$ in the simulated wakefields that correspond to the FREM images in (d--f). The color scale in (a--f) shows the relative intensity of the signal on the screen where $0$ corresponds to the intensity of the unperturbed probe beam.
        The vertical lines in (d--i) show the plasma period corresponding to $n_0 = \SI{5e17}{cm^{-3}}$. The red color in (g--i) shows the intensity of the axiparabola laser field.
        }
\end{figure*}

The structured-light wakefield was probed at different locations along the focal depth by steering the probe beam and compensating the time delay accordingly. Figure \ref{fig:Data} (a--c) shows three experimentally measured FREM images of the wakefield taken at different depths into the nozzle along the propagation axis. Figure \ref{fig:Data} (a) was taken around the start of the focal line, where the focused beam begins to drive a wakefield. Figure \ref{fig:Data} (b) was taken 0.5 mm further into the nozzle, and figure \ref{fig:Data} (c) is a further 0.5 mm after that, thus giving snapshots of the wakefield at different points along the focal depth. As can be seen, the structure of the wakefield evolves over the course of the focal depth. As the wakefield evolves, a pronounced V-shaped structure begins to develop around the central axis of the wakefield. In figure \ref{fig:Data} (a), as the pulse is beginning to generate a fully formed wakefield, this structure is barely discernible, growing more distinct 0.5 mm later in figure \ref{fig:Data} (b). 1 mm into the focal depth, in figure \ref{fig:Data} (c), the FREM image is dominated by this V-shaped structure. All the experimental images are cropped in order to allow a direct comparison with simulations. The full-sized experimental image can be seen in extended data figure \ref{fig:frem_raw}.

To understand the structure of the wake excited by an axiparabola-focused laser pulse, particle-in-cell (PIC) simulations of its interaction with a gas target were performed. First, the propagation of the reflected laser pulse from the axiparabola surface to the entrance to the gas jet was calculated using the Axiprop code \cite{Andriyash_Axiprop,Oubrerie_JoO_2022}.
The simulations used an on-axis axiparabola with parameters corresponding to the experiment. The obtained distribution of the laser field was then used in PIC simulations with the quasi-3D spectral code FBPIC \cite{Lehe_ComPhysCom_2016}. The laser pulse interacted with a trapezoidal density profile, and ionization effects were fully taken into account. The calculated electromagnetic field distribution in the wake was then used to simulate FREM images by propagating a probe electron bunch---with parameters similar to the parameters of the experimental probe bunch---through the wake. 

Figure \ref{fig:Data} (d--f) shows simulated FREM images at corresponding points along the focal depth. The images follow a similar evolution of the wake structure to that seen in the experiment, indicating that the simulations correctly capture the essential dynamics in this novel wakefield regime. Matching the experimental and simulated FREM images provides a more detailed view into the wakefields via the simulations. The corresponding simulated wakefields are shown in figure \ref{fig:Data} (g--i). As the pulse propagates along the focal line, radially off-axis wakes develop, in addition to the central, on-axis wake. As the focal depth increases, the wakes also shrinks in the transverse direction, following the evolution of the size of the central focal spot seen in figure \ref{fig:Setup} (b). 

One important thing that can be discerned by comparing the simulated FREM images to their wakefields is that the periodicity seen in the FREM images corresponds to the periodicity of the wakefield structure. Since the wakefield period is equal to the plasma wavelength, this allows for an \textit{in situ} measurement of the plasma wavelength and, therefore, of the plasma density \cite{Chen_Plenum_1983,Wan_2023_LSA_12_116}. This is emphasized in figure \ref{fig:Data} by the dashed lines which show that both the length of the wakefield in (g--i) and the length of the periodic structure in the FREM (d--f) correspond to the plasma wavelength.

The wavelength can be algorithmically reconstructed by analyzing the 2D Fourier transform of the FREM image.
When applied to simulated FREM images for all available simulation steps between 2 and 8\,mm inside the nozzle (see extended data figure~\ref{fig:density_reconstruction}), this method yields the mean reconstructed density of \SI{4.8e17}{cm^{-3}} with a standard deviation of \SI{4.5}{\percent}. The actual plasma density used in the PIC simulations was \SI{5e17}{cm^{-3}}, which confirms the reconstruction results.
Only two to three periods of the wake fit into the simulation box which limits the accuracy of the Fourier analysis. The actual FREM image can capture tens of plasma periods (see extended figure~\ref{fig:frem_raw}), thus the reconstruction based on experimental FREM images might have even higher accuracy.

\subsection*{Spatio-Temporal Coupling Effect on Wakefield Structure}

\begin{figure*}[t!]
		\centering
		\includegraphics[width=\linewidth]{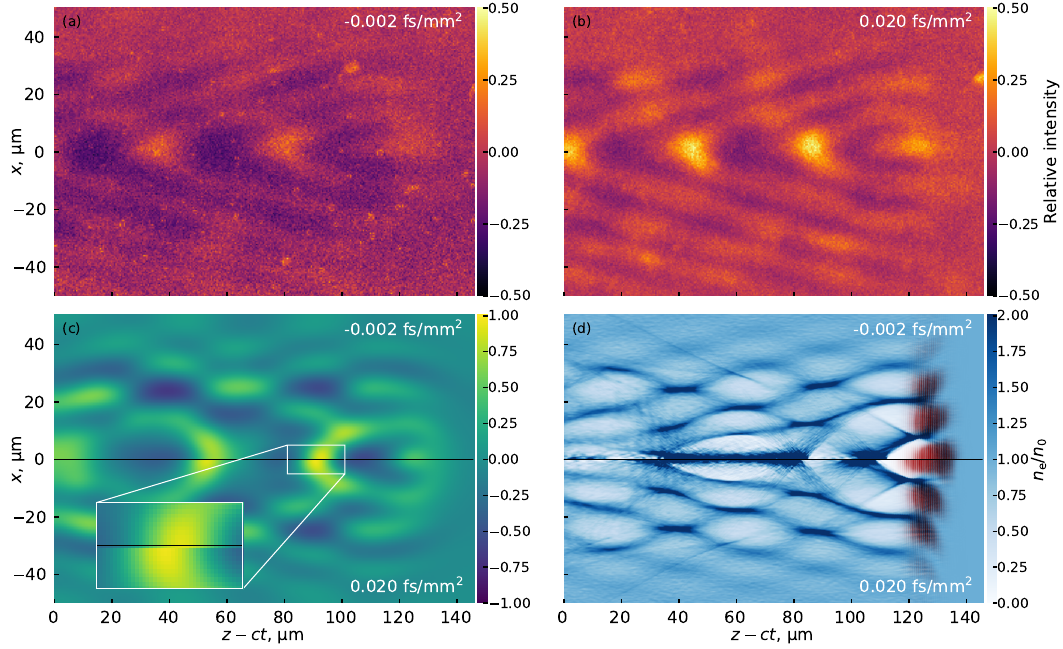}
		\caption{\label{fig:takeaway} \small (a--b) Experimentally obtained FREM images at the same point along the focal depth for PFC values of $-0.002$ and \SI{0.020}{fs/mm^2}, respectively.
        (c) Simulated FREM images for PFC values \SI{-0.002}{fs/mm^2} (top) and \SI{0.020}{fs/mm^2} (bottom) and the (d) corresponding relative electron density distribution $n_\mathrm{e}/n_0$ in PIC simulations.
        The color scale in (a--c) shows the relative intensity of the signal where $0$ is the unperturbed beam intensity. The red color in (d) shows the intensity of the axiparabola laser field. }
\end{figure*}

To investigate the nature of the prominent V-shaped structure, the relationship between the angle of this structure and the near-field spatio-temporal pulse front was examined. As was shown in Refs.~\cite{Caizergues_NaturePhotonics_2020,Palastro_PRL_2020}, radially dependent pulse delay in the near-field, when focused by an axiparabola, changes the phase velocity of the wakefield. The pulse delay can be mathematically described by a spatio-spectral phase term, $\alpha r^2 (\omega - \omega_0)$. In order to test whether the angular structure relates to the velocity of propagation of the wakefield, FREM shots were taken at different pulse-front curvatures (PFCs) of the beam to see if the effect could be seen in the FREM images. Figures \ref{fig:takeaway} (a, b) show the experimentally obtained FREM images for PFC values $\alpha$ of \SI{-0.002}{fs/mm^2} and \SI{0.020}{fs/mm^2}, respectively. Figure \ref{fig:takeaway} (c) shows the simulated FREM images for the pulses with the same PFC values (top and bottom, respectively), while figure \ref{fig:takeaway} (d) shows the corresponding plasma wakefields in PIC simulations. 

The PIC simulations allow for an absolute comparison of the relative positions of the wakefields at the same moment in time. As can be seen in the comparisons in \ref{fig:takeaway} (c) and (d), the difference in PFC induces a propagation velocity change, with the positive PFC wakefield arriving temporally later than the negative PFC wakefield.
The measured shift in the PIC simulations of the on-axis peak intensity position due to the PFC difference was equal to \SI{2.3}{\um} at this propagation distance (5.1\,mm inside the nozzle or 2.1\,mm into the focal line). This small shift does not seem to significantly affect the structure of the wake itself, as can be seen in the FREM images and the simulated wakefields. This shift, however, is critical for the use of the axiparabola for dephasingless LWFA \cite{Caizergues_NaturePhotonics_2020,Palastro_PRL_2020} as it is of the order of the correction needed for phase-locking the wakefield to the trapped electrons. 

\subsection*{The Origin of the Wakefield Structure}
\begin{figure}[t!]
		\centering
		\includegraphics[]{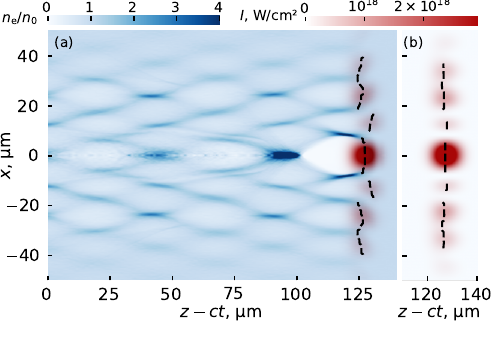}
		\caption{\label{fig:vacuum} \small Distributions of the laser field intensity of the axiparabola-focused pulse and relative electron density $n_\mathrm{e} / n_0$ in the excited wake at the propagating distance of \SI{5.5}{mm} (a) with and (b) without plasma.
        The dashed line shows the dependence of the position of the peak intensity on the transverse coordinate. }
\end{figure}

Another possibility that was examined is that the V-shaped structure is determined by the inherent transverse intensity distribution profile, which changes along the focal depth of the axiparabola.
To analyze this, the intensity distribution profiles of an axiparabola-focused pulse at a given distance with and without plasma were compared. The result is shown in figure \ref{fig:vacuum} (a) and (b), respectively. In this comparison, the intensity distribution is noticeably affected by the propagation of the pulse in plasma, which imposes tilts on the peak intensity profile.
These tilts are largely responsible for changing the relative phases of the off-axis wakefield structure seen in figure \ref{fig:vacuum} (a) which, in turn, creates the V-shaped pattern in the FREM images in figure \ref{fig:Data}.
In the absence of the tilts, the wakefield and thus the probe image would have a checkerboard pattern instead.
Therefore, the structure of the wakefield depends on the interaction of the axiparabola-focused laser with plasma and cannot be understood by looking at the laser field distribution in vacuum. This means that the direct study of the behavior in plasma is critical to understanding and optimizing this novel wakefield for use in dephasingless acceleration. To properly capture these effects, the ionization of the plasma should always be fully taken into account. 

\begin{figure}[t!]
		\centering
		\includegraphics[width=\linewidth]{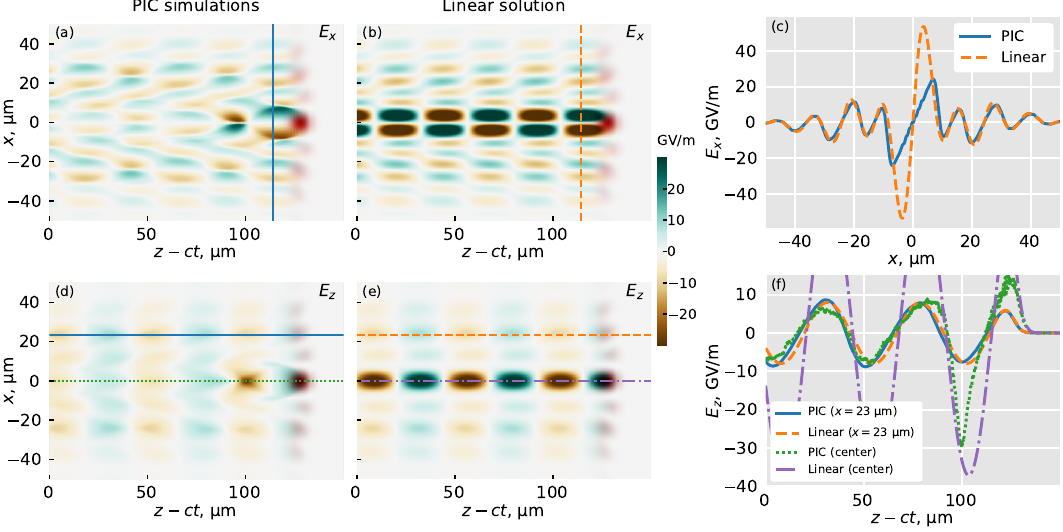}
		\caption{\label{fig:linear} \small Comparison of the spatial distribution of the transverse electric field $E_x$ between (a) the PIC simulation and (b) the calculated linear solution. (c) Transverse distribution of $E_x$ for a slice $z - ct = \SI{114}{\um}$ shown with vertical lines in (a--b). Comparison of the longitudinal electric field $E_z$ for (d) the PIC simulation and (e) the linear solution. (f) Longitudinal distributions of $E_z$ at the off-axis coordinate $x = \SI{23}{\um}$ for the PIC (solid blue) and linear (dashed orange) and at the center of the wake for the PIC (green dotted) and linear (purple dashed-dotted). The red color in (a--b, d--e) shows the distribution of the laser pulse intensity.}
\end{figure}

\subsection*{Combination of Linear and Nonlinear Wakefields}

An analysis of the electro-magnetic fields inside of the structured-light wakefield reveals another unusual property: the simultaneous mixing of linear and nonlinear wakes. The on-axis part of the excited wakefield is nonlinear, as the value of the normalized field amplitude $a_0 = e E_0/(m c \omega_\mathrm{laser})$ reaches almost 2, according to the PIC simulations. However, the off-axis wakefields are in the linear regime, with a local $a_0$ value well below 1.
This can be shown by comparing the electro-magnetic fields in the simulated wake to a linear analytical solution for the same laser distribution \cite{Gorbunov_1987_SPJETP_66_290, Sprangle_1988_APL_53_2146}. 

Figure \ref{fig:linear} (a) and (b) show the transverse ($E_x$) field component for the PIC simulated (a) and linear (b) wake, taken at the same position as the wakefield shown in figure~\ref{fig:vacuum} (a). Figure \ref{fig:linear} (c) shows a 1-D comparison of a transverse slice of the fields for the simulated (solid blue) and linear (dashed orange) wakes, at the location indicated by these lines in (a) and (b), respectively. This comparison shows that the field simultaneously exhibits behavior of both a linear and nonlinear wake. Off-axis, the simulation closely follows the linear solution. Around the central axis, however, there are typical features of a nonlinear wakefield, with a linear dependence of $E_x$ on the transverse coordinate. Around the axis, the linear solution significantly overestimates the expected field amplitude, predicting a value of around \SI{50}{GV/m}, while the simulated field is closer to \SI{20}{GV/m}. 

Figure \ref{fig:linear} (d) and (e) similarly show the longitudinal ($E_z$) field component, while figure \ref{fig:linear} (f) shows longitudinal slices of the fields. The slices are taken from the simulated field, both off-axis (solid blue) and on-axis (dotted green), as well as for the off-axis (dashed orange) and on-axis (dash-dotted purple) linear wake. The off-axis slices agree almost perfectly, both yielding a predicted field of around \SI{10}{GV/m}. The on-axis slice from the simulation, however, shows a distinctly nonlinear behavior, with a sharp negative peak at the back of the first bucket. The linear solution again overestimates the expected field amplitude, predicting a value of \SI{37}{GV/m}, while the simulated on-axis field is closer to \SI{10}{GV/m} and reaches the value of $\SI{30}{GV/m}$ only at the back of the first bucket. 

This discrepancy is explained by the difference in scaling between linear and nonlinear wakefields. In the linear regime, the field scales as $a_0^2$, while in nonlinear interactions, the scaling becomes $\propto \sqrt{a_0}$ for $a_0 \gtrsim 2$ \cite{Lu_PhysRevSTAB_2007}. 

For reference, the cold nonrelativistic wavebreaking field is $E_\mathrm{p} = m c \omega_\mathrm{p}/e \approx \SI{68}{GV/m}$ for the considered electron plasma density of \SI{5e17}{cm^{-3}}. This indicates that while the central part of the wakefield is nonlinear, the interaction does not reach the strongly nonlinear regime.

This analysis of the wakefields shows that the experimental FREM images presented in figure~\ref{fig:Data} capture both a nonlinear and a linear wake in one shot. This is very different from wakefields excited by a pulse with the same characteristics that is reflected by a regular parabolic mirror, where only one strongly-nonlinear bubble is formed (see extended data figure~\ref{fig:linear_parabola}). This difference showcases unique features of the structured-light wakefield.

\section*{Discussion}

This paper presents the first direct observation of a wakefield generated using a structured light pulse, shaped by an axiparabola and spatio-temporal couplings. As was shown, the structure of this wakefield develops over the focal depth of the axiparabola and differs significantly from that of a standard wakefield accelerator. The results were reinforced with PIC simulations which recreated the structures seen in the FREM images and yielded insight about the wakefield itself. These include the presence of radially offset wakes in addition to the central, on-axis wake; the simultaneous mixing of linear and nonlinear wakes; and the tilting of the phase front of the wake, which yields the V-shaped structure seen in the FREM images. Both the data and simulations showed that modifications to the spatio-temporal couplings of the beam, critical to tuning the propagation velocity of the wakefield, have minimal effect on the structure of the wakefield at the values relevant for dephasingless acceleration. An analysis of the periodicity of the wakefield demonstrated that the \textit{in situ} plasma density can be extracted from the FREM images, yielding a direct measurement of an important experimental parameter.

Since the FREM diagnostic does not interfere with the structured-light wakefield, the density measurement and further FREM studies can be conducted while simultaneously optimizing for electron acceleration. The results shown here and the future experimental capability of the setup could bridge the gap between the promising simulations of the novel regime of the structured-light wakefield and the lagging experimental results. With the development of minimally intrusive single-shot measurements of the spatio-temporal behavior of the pulse \cite{Smartsev_OL_2024}, further studies could be made into the effect of spatio-temporally shaping the beam on the structure and behavior of the wakefield. These insights will help facilitate the realization of the potential of the axiparabola-generated wakefield to overcome the dephasing limit of laser-wakefield acceleration. If successfully achieved and implemented in the new generation of laser facilities now coming online, such as the \SI{10}{PW} facility at ELI-NP \cite{Radier_2022_HPLSE_10}, these structured-light-based wakefields could accelerate electrons to energies in excess of 100\,GeV \cite{Caizergues_NaturePhotonics_2020}. This would constitute the highest electron energy ever accelerated. 

In addition to further studying dephasingless acceleration, the FREM technique described here can be applied to the study of other exotic wakefield structures. A notable example is the proposed use of helical beams to generate laser-wakefield accelerators capable of efficiently accelerating positrons \cite{Vieira_PRL_2014,Mendoca_PoP_2014}. The realization of such a scheme, coupled with the successful demonstration of dephasingless electron acceleration, would enable a next-generation of high-energy physics experiments based on extreme light-matter interactions.

\section*{Methods}\label{Methods}
\subsection*{Laser}

The laser system provided two temporally synchronized 2.5\,J, 27\,fs laser pulses with a central wavelength of \SI{800}{\nm}. During the experiment, the beams each delivered 1\,J on target. The two pulses are generated by splitting a 7\,J, uncompressed beam into two 3.5\,J beams which are then individually compressed, yielding two compressed 2.5 J\, beams. Therefore, they were synchronized in time, up to controllable delays from the independent beam transport lines and a femtosecond-scale jitter. Both beams had an unfocused diameter of around 50\,mm. Beam 1 was focused by a 480\,mm nominal focal length ($f/9.6$) off-axis axiparabola with a focal depth, $\delta$, of 5\,mm, and an off-axis angle of 10 degrees. The axiparabola focus had the following functional form: $f(r) = f_0 + \delta (r/R)^2$, where $f_0$ is the nominal focal length, $r$ is the radial distance from the central axis, and $R$ is the full radius of the beam. Beam 1 was focused onto a \SI{4}{\mm} wide and \SI{15}{\mm} long supersonic slit nozzle, shown in figure \ref{fig:Setup} as ``Jet 1''. The gas used was a mixture of 97\% helium and 3\% nitrogen. The plasma density in Jet 1 was around $\SI{5e17}{\cm^{-3}}$. Beam 2 was focused by a 1.5\,m focal length ($f/30$), 11 degree off-axis parabolic mirror into a gas jet generated by a supersonic converging-diverging nozzle with a throat diameter of 0.5\,mm and an outlet diameter of 3\,mm, shown in figure \ref{fig:Setup} as ``Jet 2''.  The same helium-nitrogen mixture was used in Jet 2. 

\subsection*{Probe Electron Bunch}

Upon focusing into Jet 2, Beam 2 generated an LWFA, accelerating electrons injected into the wakefield via ionization injection of the helium-nitrogen gas mixture. To characterize the electrons, they were initially passed through a spectrometer which utilized a 10-cm-long 1\,Tesla magnetic dipole and a Lanex scintillator. The magnetic dipole imposed an energy dependent angular deviation onto the electrons, allowing the scintillator to provide energy resolution of the electron bunch. Figure \ref{fig:B2_Lanex} shows a sample angularly resolved spectrum of the electrons accelerated by Beam 2 and used as the electrons for the FREM probe. A quasi-monoenergetic peak was observed at 270\,MeV $\pm$ 35\,MeV and the FWHM of the angular divergence was 1.6\,mrad.

\subsection*{FREM}

Jet 2 was spatially arranged such that the electrons from the LWFA in Jet 2 propagated 10\,cm before intersecting with the structured-light wakefield in Jet 1. By that point, the probe beam electrons spatially spread out to a lateral extent of several hundred micrometers, allowing them to illuminate a large section of the wakefield. The probe beam electrons were spatially and temporally overlapped with the structured-light wakefield. Rough overlap was achieved via the use of Beam 2 for shadowgraphy of the ionization front of Beam 1 in Jet 1. This overlap was sufficient to ensure synchronization on the order of a few hundred femtoseconds and spatial overlap on the order of a hundred micrometers. 

Further alignment was done using the electron probe, which impinged onto the axiparabola-generated wakefield at near normal incidence. After crossing the wakefield, the electrons were allowed to drift a further 7\,mm in vacuum before they hit a 30 micrometer thick Ce:YAG scintillator. This allowed the momentum kicks imparted by the wakefield to become density modulations which could be seen on the scintillator. The Ce:YAG was shielded from residual laser light by a stainless steel film of 100 micrometers thickness. The light emitted by the Ce:YAG screen was collected by a Mitutoyo infinity-corrected, long-working-distance, plan-apochromatic 10X microscope objective and a Thorlabs achromatic lens with a focal length of 300\,mm onto a Hamamatsu ORCA-FLASH4.0 digital CMOS camera. The imaging resolution was checked with a 1951 USAF resolution test target, giving a limit of \SI{2.2}{\um}. For the plots in figures \ref{fig:Data} and \ref{fig:takeaway}, the measured spatial dimensions were scaled down by a factor of 1.07 to account for probe beam transverse expansion after the interaction and to correspond to the actual dimensions of the probed wakefields. Experimental FREM images were also cleaned from noise and background irradiation during post-processing (see figure~\ref{fig:frem_raw}).

The plasma wavelength is reconstructed from the FREM images by applying a 2D Fourier transform to the image, taking a sum of the absolute spectral amplitude over transverse (vertical) wavenumbers, and then taking the longitudinal wavenumber corresponding to the peak value of the sum.
The densities estimated from the experimental FREM images in figures~\ref{fig:Data}(a--c) are \SI{5.9e17}{cm^{-3}}, \SI{4.5e17}{cm^{-3}}, \SI{6.0e17}{cm^{-3}}, respectively.

\subsection*{Axiprop Simulation}

To simulate the interactions of the gas with the axiparabola-focused beam, the propagation of the reflected laser pulse from the axiparabola surface to the focal plane was calculated using the Axiprop code \cite{Andriyash_Axiprop,Oubrerie_JoO_2022}.
In order to make the beam compatible with the axisymmetric solver of Axiprop and the quasi-3D spectral code FBPIC \cite{Lehe_ComPhysCom_2016}---which uses angular mode decomposition and is best suited for axially symmetric beams---an on-axis axiparabola with the parameters corresponding to the experiment (\SI{480}{\mm} focal length, \SI{5}{\mm} focal depth for \SI{25}{\mm} radius) was used.
The laser pulse had a Gaussian temporal profile with a duration of \SI{27}{\fs} (FWHM intensity) with a central wavelength of \SI{800}{\nm} and a 6th order super-Gaussian transverse profile with a diameter (FWHM intensity) of \SI{40}{\mm}.
The energy of the pulse was equal to \SI{0.6}{\J}.
The lower energy and diameter compared to the experimental setup were used for better correspondence with the experimental observations, perhaps reflecting non-ideal alignment of the axiparabola or phase-front defects of the real laser pulse as well as laser energy loses in the beam transport line.
For simulations with PFC, an additional PFC phase correction was applied to the laser pulse on the axiparabola surface.

For the PIC simulations with the parabola-reflected pulse in figure~\ref{fig:linear_parabola}, Axiprop simulations with a pulse reflected by a regular parabola with the same focal length of \SI{480}{mm} were performed.

\subsection*{Particle-in-cell Simulation}

The obtained distribution of the laser field was saved in the LASY format \cite{thevenet_lasy} and then input into PIC simulations performed with the quasi-3D spectral code FBPIC with azimuthal mode decomposition \cite{Lehe_ComPhysCom_2016}, assuming linear polarization in the $x$ direction.
To make sure the entire laser pulse fits inside the simulation box, a simulation box with a large size of \SI{586}{\um} in the transverse $r$ direction and \SI{150}{\um} in the longitudinal $z$ direction, with two azimuthal modes was used.
The grid resolution was $dz = \SI{0.04}{\um}$ and $dr = \SI{0.24}{\um}$, respectively.
To accelerate the simulation, a Lorentz-boosted frame with a Lorentz factor $\gamma = 2.5$ was used \cite{Lehe_2016_PRE_94_53305}.

The slit nozzle gas target was represented by an 8-mm plateau with 1\,mm linear up and downramps with a peak electron density at full ionization of \SI{5e17}{\cm^{-3}}.
This expected density was estimated based on the plasma wavelength retrieved from the experimental FREM images.
We used pure helium as the gas in the PIC simulations to limit the number of macroparticles and accelerate the simulation, as the small nitrogen component present in the experiments is not expected to be important in the absence of ionization injection in Jet 1.
The gas was initially neutral to properly account for the diffraction of the weaker laser field at large radii where the gas might be partially ionized.
The gas was initialized with 32 helium atoms per 2D cell (2 in $r$, 2 in $z$, 8 in $\theta$ directions, respectively).
The starting point of the axiparabola focal depth $f_0$ corresponded to a position of \SI{3}{\mm} (\SI{2}{\mm} into the plateau) in the simulations.

For the simulations of the wakefield generated by the parabola-reflected pulse in the extended data figure~\ref{fig:linear_parabola}, the same PIC simulation and gas target parameters were used with the focal point of the parabola at 3\,mm inside the nozzle. The only change was the reduction of the transverse size of the box to \SI{166}{\um}.

\subsection*{FREM Simulation}

The calculated electromagnetic field distribution in the wake was then used to simulate FREM images by propagating a probe electron bunch with parameters similar to the parameters of the experimental probe bunch through the field. The field was assumed to be moving perpendicular to the beam at the speed of light.
The simulated bunch had an energy spectrum with a uniform energy distribution between \SI{100}{\MeV} and \SI{300}{\MeV}, a duration of \SI{10}{\fs} with a rectangular current profile, an initial transverse size at the source of \SI{5}{\um}, and a divergence at maximum energy of \SI{1}{mrad}.
The bunch was modeled by \num{5e6} particles of equal weights.

Before interacting with the wakefield, the bunch propagated for \SI{10}{\cm} in vacuum.
After the interaction, it was projected to a plane \SI{7}{\mm} away from the wakefield and normal to the bunch propagation direction.
The projection used a pixel size of \SI{0.52}{\um} (equal to the experimental one).
An additional Gaussian filter with a kernel size of 8 pixels (standard deviation) was applied to the image to model the resolution constraints in the experimental imaging system as well as to reduce the excessive noise introduced by a limited number of particles in the simulations.
To eliminate the effect of varying background brightness, the signal on the plane was calculated relative to the signal in the absence of the wakefield.
Similar to the experimental FREM images, the spatial dimensions on the image plane were scaled down by a factor of 1.07 to account for the beam expansion during its propagation to the screen.

\subsection*{Linear Solution}

In a linear wakefield, the plasma dynamics can be described by linearized hydrodynamic equations, assuming a small laser pulse amplitude $a_0$ and a corresponding small perturbation of the plasma density \cite{Gorbunov_1987_SPJETP_66_290, Sprangle_1988_APL_53_2146}.
This procedure enables the calculation of the electrostatic potential
\begin{equation}
    \varphi(x, y, z-ct) = \frac{m c^2}{2e} \int \left\langle{a^2}\right\rangle(x, y, \xi') \sin[k_\mathrm{p} (z - ct - \xi')] d\xi',
\end{equation}
where $\left\langle a^2 \right\rangle$ is the time-averaged normalized laser field amplitude $a(x,y,z-ct) = e E_x/(m c \omega_\mathrm{p})$ which is retrieved from the PIC simulation data, and $k_\mathrm{p} = \omega_\mathrm{p} / c$ is the plasma wavenumber corresponding to the electron density $n_0 = \SI{5e17}{cm^{-3}}$.
The electric field is then retrieved by taking the gradient of $\varphi$.
The magnetic field in the linear wake is negligibly small ($\propto a_0^4$).

\backmatter








\section*{Declarations}

\begin{itemize}
\item Funding: The research was supported by the Schwartz/Reisman Center for Intense Laser Physics, the Benoziyo Endowment Fund for the Advancement of Science, the Israel Science Foundation, Minerva, Wolfson Foundation, the Schilling Foundation, R. Lapon, Dita and Yehuda Bronicki, and the Helmholtz Association.  
\item Competing interests: The authors declare no competing interests.
\item Ethics approval and consent to participate: Not Applicable 
\item Consent for publication: Not Applicable
\item Data availability: Available upon reasonable request to the authors. 
\item Materials availability: Not applicable
\item Code availability: Available upon reasonable request to the authors. 
\item Author contribution: A.L. and A.G. prepared the manuscript with the supervision of V.M. and with input from all the authors; A.L., S.S., and S.T. performed the experiments, with the support of E.Y.L., E.K., and S.B.; A.G. performed the simulations with the support of A.L. and I.A.A.; S.S. designed the axiparabola together with the spatio-temporal control doublet; All authors discussed the results and V.M. provided overall guidance to the project. 
\end{itemize}

\bibliography{paper}


\begin{thebibliography}{49}
\ifx \bisbn   \undefined \def \bisbn  #1{ISBN #1}\fi
\ifx \binits  \undefined \def \binits#1{#1}\fi
\ifx \bauthor  \undefined \def \bauthor#1{#1}\fi
\ifx \batitle  \undefined \def \batitle#1{#1}\fi
\ifx \bjtitle  \undefined \def \bjtitle#1{#1}\fi
\ifx \bvolume  \undefined \def \bvolume#1{\textbf{#1}}\fi
\ifx \byear  \undefined \def \byear#1{#1}\fi
\ifx \bissue  \undefined \def \bissue#1{#1}\fi
\ifx \bfpage  \undefined \def \bfpage#1{#1}\fi
\ifx \blpage  \undefined \def \blpage #1{#1}\fi
\ifx \burl  \undefined \def \burl#1{\textsf{#1}}\fi
\ifx \doiurl  \undefined \def \doiurl#1{\url{https://doi.org/#1}}\fi
\ifx \betal  \undefined \def \betal{\textit{et al.}}\fi
\ifx \binstitute  \undefined \def \binstitute#1{#1}\fi
\ifx \binstitutionaled  \undefined \def \binstitutionaled#1{#1}\fi
\ifx \bctitle  \undefined \def \bctitle#1{#1}\fi
\ifx \beditor  \undefined \def \beditor#1{#1}\fi
\ifx \bpublisher  \undefined \def \bpublisher#1{#1}\fi
\ifx \bbtitle  \undefined \def \bbtitle#1{#1}\fi
\ifx \bedition  \undefined \def \bedition#1{#1}\fi
\ifx \bseriesno  \undefined \def \bseriesno#1{#1}\fi
\ifx \blocation  \undefined \def \blocation#1{#1}\fi
\ifx \bsertitle  \undefined \def \bsertitle#1{#1}\fi
\ifx \bsnm \undefined \def \bsnm#1{#1}\fi
\ifx \bsuffix \undefined \def \bsuffix#1{#1}\fi
\ifx \bparticle \undefined \def \bparticle#1{#1}\fi
\ifx \barticle \undefined \def \barticle#1{#1}\fi
\bibcommenthead
\ifx \bconfdate \undefined \def \bconfdate #1{#1}\fi
\ifx \botherref \undefined \def \botherref #1{#1}\fi
\ifx \url \undefined \def \url#1{\textsf{#1}}\fi
\ifx \bchapter \undefined \def \bchapter#1{#1}\fi
\ifx \bbook \undefined \def \bbook#1{#1}\fi
\ifx \bcomment \undefined \def \bcomment#1{#1}\fi
\ifx \oauthor \undefined \def \oauthor#1{#1}\fi
\ifx \citeauthoryear \undefined \def \citeauthoryear#1{#1}\fi
\ifx \endbibitem  \undefined \def \endbibitem {}\fi
\ifx \bconflocation  \undefined \def \bconflocation#1{#1}\fi
\ifx \arxivurl  \undefined \def \arxivurl#1{\textsf{#1}}\fi
\csname PreBibitemsHook\endcsname

\bibitem[\protect\citeauthoryear{Tajima and Dawson}{1979}]{Tajima_PRL_1979}
\begin{barticle}
\bauthor{\bsnm{Tajima}, \binits{T.}},
\bauthor{\bsnm{Dawson}, \binits{J.M.}}:
\batitle{Laser electron accelerator}.
\bjtitle{Physical Review Letters}
\bvolume{43}(\bissue{4}),
\bfpage{267}--\blpage{270}
(\byear{1979})
\doiurl{10.1103/PhysRevLett.43.267}
\end{barticle}
\endbibitem

\bibitem[\protect\citeauthoryear{Mangles et~al.}{2004}]{Mangles_Nature_2004}
\begin{barticle}
\bauthor{\bsnm{Mangles}, \binits{S.P.D.}},
\bauthor{\bsnm{Murphy}, \binits{C.D.}},
\bauthor{\bsnm{Najmudin}, \binits{Z.}},
\bauthor{\bsnm{Thomas}, \binits{A.G.R.}},
\bauthor{\bsnm{Collier}, \binits{J.L.}},
\bauthor{\bsnm{Dangor}, \binits{A.E.}},
\bauthor{\bsnm{Divall}, \binits{E.J.}},
\bauthor{\bsnm{Foster}, \binits{P.S.}},
\bauthor{\bsnm{Gallacher}, \binits{J.G.}},
\bauthor{\bsnm{Hooker}, \binits{C.J.}},
\bauthor{\bsnm{Jaroszynski}, \binits{D.A.}},
\bauthor{\bsnm{Langley}, \binits{A.J.}},
\bauthor{\bsnm{Mori}, \binits{W.B.}},
\bauthor{\bsnm{Norreys}, \binits{P.A.}},
\bauthor{\bsnm{Tsung}, \binits{F.S.}},
\bauthor{\bsnm{Viskup}, \binits{R.}},
\bauthor{\bsnm{Walton}, \binits{B.R.}},
\bauthor{\bsnm{Krushelnick}, \binits{K.}}:
\batitle{Monoenergetic beams of relativistic electrons from intense
  laser–plasma interactions}.
\bjtitle{Nature}
\bvolume{431},
\bfpage{535}--\blpage{538}
(\byear{2004})
\doiurl{10.1038/nature02939}
\end{barticle}
\endbibitem

\bibitem[\protect\citeauthoryear{Geddes et~al.}{2004}]{Geddes_Nature_2004}
\begin{barticle}
\bauthor{\bsnm{Geddes}, \binits{C.G.R.}},
\bauthor{\bsnm{Toth}, \binits{C.}},
\bauthor{\bsnm{Tilborg}, \binits{J.}},
\bauthor{\bsnm{Esarey}, \binits{E.}},
\bauthor{\bsnm{Schroeder}, \binits{C.B.}},
\bauthor{\bsnm{Bruhwiler}, \binits{D.}},
\bauthor{\bsnm{Nieter}, \binits{C.}},
\bauthor{\bsnm{Cary}, \binits{J.}},
\bauthor{\bsnm{Leemans}, \binits{W.P.}}:
\batitle{High-quality electron beams from a laser wakefield accelerator using
  plasma-channel guiding}.
\bjtitle{Nature}
\bvolume{431},
\bfpage{538}--\blpage{541}
(\byear{2004})
\doiurl{10.1038/nature02900}
\end{barticle}
\endbibitem

\bibitem[\protect\citeauthoryear{Faure et~al.}{2004}]{Faure_Nature_2004}
\begin{barticle}
\bauthor{\bsnm{Faure}, \binits{J.}},
\bauthor{\bsnm{Glinec}, \binits{Y.}},
\bauthor{\bsnm{Pukhov}, \binits{A.}},
\bauthor{\bsnm{Kiselev}, \binits{S.}},
\bauthor{\bsnm{Gordienko}, \binits{S.}},
\bauthor{\bsnm{Lefebvre}, \binits{E.}},
\bauthor{\bsnm{Rousseau}, \binits{J.-P.}},
\bauthor{\bsnm{Burgy}, \binits{F.}},
\bauthor{\bsnm{Malka}, \binits{V.}}:
\batitle{A laser–plasma accelerator producing monoenergetic electron beams}.
\bjtitle{Nature}
\bvolume{431},
\bfpage{541}--\blpage{544}
(\byear{2004})
\doiurl{10.1038/nature02963}
\end{barticle}
\endbibitem

\bibitem[\protect\citeauthoryear{Glinec et~al.}{2006}]{Glinec_Med_Phys_2006}
\begin{barticle}
\bauthor{\bsnm{Glinec}, \binits{Y.}},
\bauthor{\bsnm{Faure}, \binits{J.}},
\bauthor{\bsnm{Malka}, \binits{V.}},
\bauthor{\bsnm{Fuchs}, \binits{T.}},
\bauthor{\bsnm{Szymanowski}, \binits{H.}},
\bauthor{\bsnm{Oelfke}, \binits{U.}}:
\batitle{Radiotherapy with laser-plasma accelerators: Monte {C}arlo simulation
  of dose deposited by an experimental quasimonoenergetic electron beam}.
\bjtitle{Medical Physics}
\bvolume{33},
\bfpage{155}--\blpage{162}
(\byear{2006})
\doiurl{10.1118/1.2140115}
\end{barticle}
\endbibitem

\bibitem[\protect\citeauthoryear{Glinec et~al.}{2005}]{Glinec_PRL_2005}
\begin{barticle}
\bauthor{\bsnm{Glinec}, \binits{Y.}},
\bauthor{\bsnm{Faure}, \binits{J.}},
\bauthor{\bsnm{Dain}, \binits{L.L.}},
\bauthor{\bsnm{Darbon}, \binits{S.}},
\bauthor{\bsnm{Hosokai}, \binits{T.}},
\bauthor{\bsnm{Santos}, \binits{J.J.}},
\bauthor{\bsnm{Lefebvre}, \binits{E.}},
\bauthor{\bsnm{Rousseau}, \binits{J.P.}},
\bauthor{\bsnm{Burgy}, \binits{F.}},
\bauthor{\bsnm{Mercier}, \binits{B.}},
\bauthor{\bsnm{Malka}, \binits{V.}}:
\batitle{High-resolution $\ensuremath{\gamma}$-ray radiography produced by a
  laser-plasma driven electron source}.
\bjtitle{Physical Review Letters}
\bvolume{94},
\bfpage{025003}
(\byear{2005})
\doiurl{10.1103/PhysRevLett.94.025003}
\end{barticle}
\endbibitem

\bibitem[\protect\citeauthoryear{Wang et~al.}{2021}]{Wang_Nature_2021}
\begin{barticle}
\bauthor{\bsnm{Wang}, \binits{W.}},
\bauthor{\bsnm{Feng}, \binits{K.}},
\bauthor{\bsnm{Ke}, \binits{L.}},
\bauthor{\bsnm{Yu}, \binits{C.}},
\bauthor{\bsnm{Xu}, \binits{Y.}},
\bauthor{\bsnm{Qi}, \binits{R.}},
\bauthor{\bsnm{Chen}, \binits{Y.}},
\bauthor{\bsnm{Qin}, \binits{Z.}},
\bauthor{\bsnm{Zhang}, \binits{Z.}},
\bauthor{\bsnm{Fang}, \binits{M.}},
\bauthor{\bsnm{Liu}, \binits{J.}},
\bauthor{\bsnm{Jiang}, \binits{K.}},
\bauthor{\bsnm{Hao~Wang}, \binits{C.W.}},
\bauthor{\bsnm{Yang}, \binits{X.}},
\bauthor{\bsnm{Wu}, \binits{F.}},
\bauthor{\bsnm{Leng}, \binits{Y.}},
\bauthor{\bsnm{Liu}, \binits{J.}},
\bauthor{\bsnm{Li}, \binits{R.}},
\bauthor{\bsnm{Xu}, \binits{Z.}}:
\batitle{Free-electron lasing at 27 nanometres based on a laser wakefield
  accelerator}.
\bjtitle{Nature}
\bvolume{595},
\bfpage{516}--\blpage{520}
(\byear{2021})
\doiurl{10.1038/s41586-021-03678-x}
\end{barticle}
\endbibitem

\bibitem[\protect\citeauthoryear{Labat
  et~al.}{2022}]{Labat_NaturePhotonics_2022}
\begin{barticle}
\bauthor{\bsnm{Labat}, \binits{M.}},
\bauthor{\bsnm{Cabadağ}, \binits{J.C.}},
\bauthor{\bsnm{Ghaith}, \binits{A.}},
\bauthor{\bsnm{Irman}, \binits{A.}},
\bauthor{\bsnm{Berlioux}, \binits{A.}},
\bauthor{\bsnm{Berteaud}, \binits{P.}},
\bauthor{\bsnm{Blache}, \binits{F.}},
\bauthor{\bsnm{Bock}, \binits{S.}},
\bauthor{\bsnm{Bouvet}, \binits{F.}},
\bauthor{\bsnm{Briquez}, \binits{F.}},
\bauthor{\bsnm{Chang}, \binits{Y.-Y.}},
\bauthor{\bsnm{Corde}, \binits{S.}},
\bauthor{\bsnm{Debus}, \binits{A.}},
\bauthor{\bsnm{Oliveira}, \binits{C.D.}},
\bauthor{\bsnm{Duval}, \binits{J.-P.}},
\bauthor{\bsnm{Dietrich}, \binits{Y.}},
\bauthor{\bsnm{Ajjouri}, \binits{M.E.}},
\bauthor{\bsnm{Eisenmann}, \binits{C.}},
\bauthor{\bsnm{Gautier}, \binits{J.}},
\bauthor{\bsnm{Gebhardt}, \binits{R.}},
\bauthor{\bsnm{Grams}, \binits{S.}},
\bauthor{\bsnm{Helbig}, \binits{U.}},
\bauthor{\bsnm{Herbeaux}, \binits{C.}},
\bauthor{\bsnm{Hubert}, \binits{N.}},
\bauthor{\bsnm{Kitegi}, \binits{C.}},
\bauthor{\bsnm{Kononenko}, \binits{O.}},
\bauthor{\bsnm{Kuntzsch}, \binits{M.}},
\bauthor{\bsnm{LaBerge}, \binits{M.}},
\bauthor{\bsnm{Lê}, \binits{S.}},
\bauthor{\bsnm{Leluan}, \binits{B.}},
\bauthor{\bsnm{Loulergue}, \binits{A.}},
\bauthor{\bsnm{Malka}, \binits{V.}},
\bauthor{\bsnm{Marteau}, \binits{F.}},
\bauthor{\bsnm{Guyen}, \binits{M.H.N.}},
\bauthor{\bsnm{Oumbarek-Espinos}, \binits{D.}},
\bauthor{\bsnm{Pausch}, \binits{R.}},
\bauthor{\bsnm{Pereira}, \binits{D.}},
\bauthor{\bsnm{Püschel}, \binits{T.}},
\bauthor{\bsnm{Ricaud}, \binits{J.-P.}},
\bauthor{\bsnm{Rommeluere}, \binits{P.}},
\bauthor{\bsnm{Roussel}, \binits{E.}},
\bauthor{\bsnm{Rousseau}, \binits{P.}},
\bauthor{\bsnm{Schöbel}, \binits{S.}},
\bauthor{\bsnm{Sebdaoui}, \binits{M.}},
\bauthor{\bsnm{Steiniger}, \binits{K.}},
\bauthor{\bsnm{Tavakoli}, \binits{K.}},
\bauthor{\bsnm{Thaury}, \binits{C.}},
\bauthor{\bsnm{Ufer}, \binits{P.}},
\bauthor{\bsnm{Valléau}, \binits{M.}},
\bauthor{\bsnm{Vandenberghe}, \binits{M.}},
\bauthor{\bsnm{Vétéran}, \binits{J.}},
\bauthor{\bsnm{Schramm}, \binits{U.}},
\bauthor{\bsnm{Couprie}, \binits{M.-E.}}:
\batitle{Seeded free-electron laser driven by a compact laser plasma
  accelerator}.
\bjtitle{Nature Photonics}
\bvolume{17},
\bfpage{150}--\blpage{156}
(\byear{2022})
\doiurl{10.1038/s41566-022-01104-w}
\end{barticle}
\endbibitem

\bibitem[\protect\citeauthoryear{Esarey
  et~al.}{2009}]{Esarey_ReviewOfModernPhysics_2009}
\begin{barticle}
\bauthor{\bsnm{Esarey}, \binits{E.}},
\bauthor{\bsnm{Schroeder}, \binits{C.B.}},
\bauthor{\bsnm{Leemans}, \binits{W.P.}}:
\batitle{Physics of laser-driven plasma-based electron accelerators}.
\bjtitle{Reviews of Modern Physics}
\bvolume{81},
\bfpage{1229}--\blpage{1285}
(\byear{2009})
\doiurl{10.1103/RevModPhys.81.1229}
\end{barticle}
\endbibitem

\bibitem[\protect\citeauthoryear{Joshi et~al.}{1984}]{Joshi_Nature_1984}
\begin{barticle}
\bauthor{\bsnm{Joshi}, \binits{C.}},
\bauthor{\bsnm{Mori}, \binits{W.B.}},
\bauthor{\bsnm{Katsouleas}, \binits{T.}},
\bauthor{\bsnm{Dawson}, \binits{J.M.}},
\bauthor{\bsnm{Kindel}, \binits{J.M.}},
\bauthor{\bsnm{Forslund}, \binits{D.W.}}:
\batitle{Ultrahigh gradient particle acceleration by intense laser-driven
  plasma density waves}.
\bjtitle{Nature}
\bvolume{311},
\bfpage{525}--\blpage{529}
(\byear{1984})
\doiurl{10.1038/311525a0}
\end{barticle}
\endbibitem

\bibitem[\protect\citeauthoryear{Sprangle et~al.}{2001}]{Sprangle_PRE_2001}
\begin{barticle}
\bauthor{\bsnm{Sprangle}, \binits{P.}},
\bauthor{\bsnm{Hafizi}, \binits{B.}},
\bauthor{\bsnm{Pe\~nano}, \binits{J.R.}},
\bauthor{\bsnm{Hubbard}, \binits{R.F.}},
\bauthor{\bsnm{Ting}, \binits{A.}},
\bauthor{\bsnm{Moore}, \binits{C.I.}},
\bauthor{\bsnm{Gordon}, \binits{D.F.}},
\bauthor{\bsnm{Zigler}, \binits{A.}},
\bauthor{\bsnm{Kaganovich}, \binits{D.}},
\bauthor{\bsnm{Antonsen}, \binits{T.M.}}:
\batitle{Wakefield generation and {GeV} acceleration in tapered plasma
  channels}.
\bjtitle{Physical Review E}
\bvolume{63},
\bfpage{056405}
(\byear{2001})
\doiurl{10.1103/PhysRevE.63.056405}
\end{barticle}
\endbibitem

\bibitem[\protect\citeauthoryear{Guillaume et~al.}{2015}]{Guillaume_PRL_2015}
\begin{barticle}
\bauthor{\bsnm{Guillaume}, \binits{E.}},
\bauthor{\bsnm{D\"opp}, \binits{A.}},
\bauthor{\bsnm{Thaury}, \binits{C.}},
\bauthor{\bsnm{Ta~Phuoc}, \binits{K.}},
\bauthor{\bsnm{Lifschitz}, \binits{A.}},
\bauthor{\bsnm{Grittani}, \binits{G.}},
\bauthor{\bsnm{Goddet}, \binits{J.-P.}},
\bauthor{\bsnm{Tafzi}, \binits{A.}},
\bauthor{\bsnm{Chou}, \binits{S.W.}},
\bauthor{\bsnm{Veisz}, \binits{L.}},
\bauthor{\bsnm{Malka}, \binits{V.}}:
\batitle{Electron rephasing in a laser-wakefield accelerator}.
\bjtitle{Physical Review Letters}
\bvolume{115},
\bfpage{155002}
(\byear{2015})
\doiurl{10.1103/PhysRevLett.115.155002}
\end{barticle}
\endbibitem

\bibitem[\protect\citeauthoryear{Leemans and
  Esarey}{2009}]{Leemans_PhysicsToday_2009}
\begin{barticle}
\bauthor{\bsnm{Leemans}, \binits{W.}},
\bauthor{\bsnm{Esarey}, \binits{E.}}:
\batitle{Laser-driven plasma-wave electron accelerators}.
\bjtitle{Physics Today}
\bvolume{62}(\bissue{3}),
\bfpage{44}--\blpage{49}
(\byear{2009})
\doiurl{10.1063/1.3099645}
\end{barticle}
\endbibitem

\bibitem[\protect\citeauthoryear{Leemans
  et~al.}{2006}]{Leemans_NaturePhysics_2006}
\begin{barticle}
\bauthor{\bsnm{Leemans}, \binits{W.P.}},
\bauthor{\bsnm{Nagler}, \binits{B.}},
\bauthor{\bsnm{Gonsalves}, \binits{A.J.}},
\bauthor{\bsnm{Toth}, \binits{C.}},
\bauthor{\bsnm{Nakamura}, \binits{K.}},
\bauthor{\bsnm{Geddes}, \binits{C.G.R.}},
\bauthor{\bsnm{Esarey}, \binits{E.}},
\bauthor{\bsnm{Schroeder}, \binits{C.B.}},
\bauthor{\bsnm{Hooker}, \binits{S.M.}}:
\batitle{{GeV} electron beams from a centimetre-scale accelerator}.
\bjtitle{Nature Physics}
\bvolume{2},
\bfpage{696}--\blpage{699}
(\byear{2006})
\doiurl{10.1038/nphys418}
\end{barticle}
\endbibitem

\bibitem[\protect\citeauthoryear{Gonsalves et~al.}{2019}]{Gonsalves_PRL_2019}
\begin{barticle}
\bauthor{\bsnm{Gonsalves}, \binits{A.J.}},
\bauthor{\bsnm{Nakamura}, \binits{K.}},
\bauthor{\bsnm{Daniels}, \binits{J.}},
\bauthor{\bsnm{Benedetti}, \binits{C.}},
\bauthor{\bsnm{Pieronek}, \binits{C.}},
\bauthor{\bsnm{Raadt}, \binits{T.C.H.}},
\bauthor{\bsnm{Steinke}, \binits{S.}},
\bauthor{\bsnm{Bin}, \binits{J.H.}},
\bauthor{\bsnm{Bulanov}, \binits{S.S.}},
\bauthor{\bsnm{Tilborg}, \binits{J.}},
\bauthor{\bsnm{Geddes}, \binits{C.G.R.}},
\bauthor{\bsnm{Schroeder}, \binits{C.B.}},
\bauthor{\bsnm{T\'oth}, \binits{C.}},
\bauthor{\bsnm{Esarey}, \binits{E.}},
\bauthor{\bsnm{Swanson}, \binits{K.}},
\bauthor{\bsnm{Fan-Chiang}, \binits{L.}},
\bauthor{\bsnm{Bagdasarov}, \binits{G.}},
\bauthor{\bsnm{Bobrova}, \binits{N.}},
\bauthor{\bsnm{Gasilov}, \binits{V.}},
\bauthor{\bsnm{Korn}, \binits{G.}},
\bauthor{\bsnm{Sasorov}, \binits{P.}},
\bauthor{\bsnm{Leemans}, \binits{W.P.}}:
\batitle{Petawatt laser guiding and electron beam acceleration to 8 {GeV} in a
  laser-heated capillary discharge waveguide}.
\bjtitle{Physical Review Letters}
\bvolume{122},
\bfpage{084801}
(\byear{2019})
\doiurl{10.1103/PhysRevLett.122.084801}
\end{barticle}
\endbibitem

\bibitem[\protect\citeauthoryear{Sprangle
  et~al.}{1988}]{Sprangle_1988_APL_53_2146}
\begin{barticle}
\bauthor{\bsnm{Sprangle}, \binits{P.}},
\bauthor{\bsnm{Esarey}, \binits{E.}},
\bauthor{\bsnm{Ting}, \binits{A.}},
\bauthor{\bsnm{Joyce}, \binits{G.}}:
\batitle{Laser wakefield acceleration and relativistic optical guiding}.
\bjtitle{Applied Physics Letters}
\bvolume{53}(\bissue{22}),
\bfpage{2146}--\blpage{2148}
(\byear{1988})
\doiurl{10.1063/1.100300}
\end{barticle}
\endbibitem

\bibitem[\protect\citeauthoryear{Debus et~al.}{2019}]{Debus_PRX_2019}
\begin{barticle}
\bauthor{\bsnm{Debus}, \binits{A.}},
\bauthor{\bsnm{Pausch}, \binits{R.}},
\bauthor{\bsnm{Huebl}, \binits{A.}},
\bauthor{\bsnm{Steiniger}, \binits{K.}},
\bauthor{\bsnm{Widera}, \binits{R.}},
\bauthor{\bsnm{Cowan}, \binits{T.E.}},
\bauthor{\bsnm{Schramm}, \binits{U.}},
\bauthor{\bsnm{Bussmann}, \binits{M.}}:
\batitle{Circumventing the dephasing and depletion limits of laser-wakefield
  acceleration}.
\bjtitle{Physical Review X}
\bvolume{9},
\bfpage{031044}
(\byear{2019})
\doiurl{10.1103/PhysRevX.9.031044}
\end{barticle}
\endbibitem

\bibitem[\protect\citeauthoryear{Sainte-Marie
  et~al.}{2017}]{Sainte-Marie_Optica_2017}
\begin{barticle}
\bauthor{\bsnm{Sainte-Marie}, \binits{A.}},
\bauthor{\bsnm{Gobert}, \binits{O.}},
\bauthor{\bsnm{Quere}, \binits{F.}}:
\batitle{Controlling the velocity of ultrashort light pulses in vacuum through
  spatio-temporal couplings}.
\bjtitle{Optica}
\bvolume{4}(\bissue{10}),
\bfpage{1298}--\blpage{1304}
(\byear{2017})
\doiurl{10.1364/OPTICA.4.001298}
\end{barticle}
\endbibitem

\bibitem[\protect\citeauthoryear{Froula
  et~al.}{2018}]{Froula_NaturePhotonics_2018}
\begin{barticle}
\bauthor{\bsnm{Froula}, \binits{D.H.}},
\bauthor{\bsnm{Turnbull}, \binits{D.}},
\bauthor{\bsnm{Davies}, \binits{A.S.}},
\bauthor{\bsnm{Kessler}, \binits{T.J.}},
\bauthor{\bsnm{Haberberger}, \binits{D.}},
\bauthor{\bsnm{Palastro}, \binits{J.P.}},
\bauthor{\bsnm{Bahk}, \binits{S.-W.}},
\bauthor{\bsnm{Begishev}, \binits{I.A.}},
\bauthor{\bsnm{Boni}, \binits{R.}},
\bauthor{\bsnm{Bucht}, \binits{S.}},
\bauthor{\bsnm{Katz}, \binits{J.}},
\bauthor{\bsnm{Shaw}, \binits{J.L.}}:
\batitle{Spatiotemporal control of laser intensity}.
\bjtitle{Nature Photonics}
\bvolume{12},
\bfpage{262}--\blpage{265}
(\byear{2018})
\doiurl{10.1038/s41566-018-0121-8}
\end{barticle}
\endbibitem

\bibitem[\protect\citeauthoryear{Caizergues
  et~al.}{2020}]{Caizergues_NaturePhotonics_2020}
\begin{barticle}
\bauthor{\bsnm{Caizergues}, \binits{C.}},
\bauthor{\bsnm{Smartsev}, \binits{S.}},
\bauthor{\bsnm{Malka}, \binits{V.}},
\bauthor{\bsnm{Thaury}, \binits{C.}}:
\batitle{Phase-locked laser-wakefield electron acceleration}.
\bjtitle{Nature Photonics}
\bvolume{14},
\bfpage{475}--\blpage{479}
(\byear{2020})
\doiurl{10.1038/s41566-020-0657-2}
\end{barticle}
\endbibitem

\bibitem[\protect\citeauthoryear{Palastro et~al.}{2020}]{Palastro_PRL_2020}
\begin{barticle}
\bauthor{\bsnm{Palastro}, \binits{J.P.}},
\bauthor{\bsnm{Shaw}, \binits{J.L.}},
\bauthor{\bsnm{Ramsey}, \binits{D.}},
\bauthor{\bsnm{Simpson}, \binits{T.T.}},
\bauthor{\bsnm{Froula}, \binits{D.H.}}:
\batitle{Dephasingless laser wakefield acceleration}.
\bjtitle{Physical Review Letters}
\bvolume{124},
\bfpage{134802}
(\byear{2020})
\doiurl{10.1103/PhysRevLett.124.134802}
\end{barticle}
\endbibitem

\bibitem[\protect\citeauthoryear{Ambat et~al.}{2023}]{Ambat_OpticsExpress_2023}
\begin{barticle}
\bauthor{\bsnm{Ambat}, \binits{M.V.}},
\bauthor{\bsnm{Shaw}, \binits{J.L.}},
\bauthor{\bsnm{Pigeon}, \binits{J.J.}},
\bauthor{\bsnm{Miller}, \binits{K.G.}},
\bauthor{\bsnm{Simpson}, \binits{T.T.}},
\bauthor{\bsnm{Froula}, \binits{D.H.}},
\bauthor{\bsnm{Palastro}, \binits{J.P.}}:
\batitle{Programmable-trajectory ultrafast flying focus pulses}.
\bjtitle{Optics Express}
\bvolume{31}(\bissue{19}),
\bfpage{31354}--\blpage{31368}
(\byear{2023})
\doiurl{10.1364/OE.499839}
\end{barticle}
\endbibitem

\bibitem[\protect\citeauthoryear{Miller
  et~al.}{2023}]{Miller_ScientificReports_2023}
\begin{barticle}
\bauthor{\bsnm{Miller}, \binits{K.G.}},
\bauthor{\bsnm{Pierce}, \binits{J.R.}},
\bauthor{\bsnm{Ambat}, \binits{M.V.}},
\bauthor{\bsnm{Shaw}, \binits{J.L.}},
\bauthor{\bsnm{Weichman}, \binits{K.}},
\bauthor{\bsnm{Mori}, \binits{W.B.}},
\bauthor{\bsnm{Froula}, \binits{D.H.}},
\bauthor{\bsnm{Palastro}, \binits{J.P.}}:
\batitle{Dephasingless laser wakefield acceleration in the bubble regime}.
\bjtitle{Scientific Reports}
\bvolume{13},
\bfpage{21306}
(\byear{2023})
\doiurl{10.1038/s41598-023-48249-4}
\end{barticle}
\endbibitem

\bibitem[\protect\citeauthoryear{Liberman et~al.}{2024}]{Liberman_OL_2024}
\begin{barticle}
\bauthor{\bsnm{Liberman}, \binits{A.}},
\bauthor{\bsnm{Lahaye}, \binits{R.}},
\bauthor{\bsnm{Smartsev}, \binits{S.}},
\bauthor{\bsnm{Tata}, \binits{S.}},
\bauthor{\bsnm{Benracassa}, \binits{S.}},
\bauthor{\bsnm{Golovanov}, \binits{A.}},
\bauthor{\bsnm{Levine}, \binits{E.}},
\bauthor{\bsnm{Thaury}, \binits{C.}},
\bauthor{\bsnm{Malka}, \binits{V.}}:
\batitle{Use of spatiotemporal couplings and an axiparabola to control the
  velocity of peak intensity}.
\bjtitle{Optics Letters}
\bvolume{49}(\bissue{4}),
\bfpage{814}--\blpage{817}
(\byear{2024})
\doiurl{10.1364/OL.507713}
\end{barticle}
\endbibitem

\bibitem[\protect\citeauthoryear{Pigeon et~al.}{2024}]{Pigeon_OE_2024}
\begin{barticle}
\bauthor{\bsnm{Pigeon}, \binits{J.J.}},
\bauthor{\bsnm{Franke}, \binits{P.}},
\bauthor{\bsnm{Chong}, \binits{M.L.P.}},
\bauthor{\bsnm{Katz}, \binits{J.}},
\bauthor{\bsnm{Boni}, \binits{R.}},
\bauthor{\bsnm{Dorrer}, \binits{C.}},
\bauthor{\bsnm{Palastro}, \binits{J.P.}},
\bauthor{\bsnm{Froula}, \binits{D.H.}}:
\batitle{Ultrabroadband flying-focus using an axiparabola-echelon pair}.
\bjtitle{Optics Express}
\bvolume{32}(\bissue{1}),
\bfpage{576}--\blpage{585}
(\byear{2024})
\doiurl{10.1364/OE.506112}
\end{barticle}
\endbibitem

\bibitem[\protect\citeauthoryear{Smartsev
  et~al.}{2020}]{Smartsev_OpticsLetters_2019}
\begin{barticle}
\bauthor{\bsnm{Smartsev}, \binits{S.}},
\bauthor{\bsnm{Caizergues}, \binits{C.}},
\bauthor{\bsnm{Oubrerie}, \binits{K.}},
\bauthor{\bsnm{Gautier}, \binits{J.}},
\bauthor{\bsnm{Goddet}, \binits{J.-P.}},
\bauthor{\bsnm{Tafzi}, \binits{A.}},
\bauthor{\bsnm{Phouc}, \binits{K.T.}},
\bauthor{\bsnm{Malka}, \binits{V.}},
\bauthor{\bsnm{Thaury}, \binits{C.}}:
\batitle{Axiparabola: a long-focal-depth, high-resolution mirror for broadband
  high-intensity lasers}.
\bjtitle{Optics Letters}
\bvolume{44}(\bissue{14}),
\bfpage{3414}--\blpage{3417}
(\byear{2020})
\doiurl{10.1364/OL.44.003414}
\end{barticle}
\endbibitem

\bibitem[\protect\citeauthoryear{Oubrerie et~al.}{2022}]{Oubrerie_JoO_2022}
\begin{barticle}
\bauthor{\bsnm{Oubrerie}, \binits{K.}},
\bauthor{\bsnm{Andriyash}, \binits{I.A.}},
\bauthor{\bsnm{Lahaye}, \binits{R.}},
\bauthor{\bsnm{Smartsev}, \binits{S.}},
\bauthor{\bsnm{Malka}, \binits{V.}},
\bauthor{\bsnm{Thaury}, \binits{C.}}:
\batitle{Axiparabola: a new tool for high-intensity optics}.
\bjtitle{Journal of Optics}
\bvolume{24}(\bissue{4}),
\bfpage{045503}
(\byear{2022})
\doiurl{10.1088/2040-8986/ac57d2}
\end{barticle}
\endbibitem

\bibitem[\protect\citeauthoryear{Geng et~al.}{2022}]{Geng_PoP_2022}
\begin{barticle}
\bauthor{\bsnm{Geng}, \binits{P.-F.}},
\bauthor{\bsnm{Chen}, \binits{M.}},
\bauthor{\bsnm{Zhu}, \binits{X.-Z.}},
\bauthor{\bsnm{Liu}, \binits{W.-Y.}},
\bauthor{\bsnm{Sheng}, \binits{Z.-M.}},
\bauthor{\bsnm{Zhang}, \binits{J.}}:
\batitle{{Propagation of axiparabola-focused laser pulses in uniform plasmas}}.
\bjtitle{Physics of Plasmas}
\bvolume{29}(\bissue{11}),
\bfpage{112301}
(\byear{2022})
\doiurl{10.1063/5.0109643}
\end{barticle}
\endbibitem

\bibitem[\protect\citeauthoryear{Ramsey et~al.}{2023}]{Ramsey_PRA_2023}
\begin{barticle}
\bauthor{\bsnm{Ramsey}, \binits{D.}},
\bauthor{\bsnm{Di~Piazza}, \binits{A.}},
\bauthor{\bsnm{Formanek}, \binits{M.}},
\bauthor{\bsnm{Franke}, \binits{P.}},
\bauthor{\bsnm{Froula}, \binits{D.H.}},
\bauthor{\bsnm{Malaca}, \binits{B.}},
\bauthor{\bsnm{Mori}, \binits{W.B.}},
\bauthor{\bsnm{Pierce}, \binits{J.R.}},
\bauthor{\bsnm{Simpson}, \binits{T.T.}},
\bauthor{\bsnm{Vieira}, \binits{J.}},
\bauthor{\bsnm{Vranic}, \binits{M.}},
\bauthor{\bsnm{Weichman}, \binits{K.}},
\bauthor{\bsnm{Palastro}, \binits{J.P.}}:
\batitle{Exact solutions for the electromagnetic fields of a flying focus}.
\bjtitle{Physical Review A}
\bvolume{107},
\bfpage{013513}
(\byear{2023})
\doiurl{10.1103/PhysRevA.107.013513}
\end{barticle}
\endbibitem

\bibitem[\protect\citeauthoryear{Geng et~al.}{2023}]{Geng_ChinesePhys_2023}
\begin{barticle}
\bauthor{\bsnm{Geng}, \binits{P.-F.}},
\bauthor{\bsnm{Chen}, \binits{M.}},
\bauthor{\bsnm{An}, \binits{X.-Y.}},
\bauthor{\bsnm{Liu}, \binits{W.-Y.}},
\bauthor{\bsnm{Zhu}, \binits{X.-Z.}},
\bauthor{\bsnm{Li}, \binits{J.-L.}},
\bauthor{\bsnm{Li}, \binits{B.-Y.}},
\bauthor{\bsnm{Sheng}, \binits{Z.-M.}}:
\batitle{Plasma density transition-based electron injection in laser wake field
  acceleration driven by a flying focus laser}.
\bjtitle{Chinese Physics B}
\bvolume{32}(\bissue{4}),
\bfpage{044101}
(\byear{2023})
\doiurl{10.1088/1674-1056/acae79}
\end{barticle}
\endbibitem

\bibitem[\protect\citeauthoryear{Liberman et~al.}{2024}]{Liberman_CLEO_2024}
\begin{botherref}
\oauthor{\bsnm{Liberman}, \binits{A.}},
\oauthor{\bsnm{Smartsev}, \binits{S.}},
\oauthor{\bsnm{Tata}, \binits{S.}},
\oauthor{\bsnm{Golovanov}, \binits{A.}},
\oauthor{\bsnm{Benracassa}, \binits{S.}},
\oauthor{\bsnm{Andriyash}, \binits{I.}},
\oauthor{\bsnm{Lahaye}, \binits{R.}},
\oauthor{\bsnm{Levine}, \binits{E.Y.}},
\oauthor{\bsnm{Kroupp}, \binits{E.}},
\oauthor{\bsnm{Thaury}, \binits{C.}},
\oauthor{\bsnm{Malka}, \binits{V.}}:
First electrons from axiparabola-based {LWFA}.
CLEO 2024
\textbf{ATh3H.1}
(2024)
\doiurl{10.1364/CLEO_AT.2024.ATh3H.1}
\end{botherref}
\endbibitem

\bibitem[\protect\citeauthoryear{Zhang
  et~al.}{2016}]{Zhang_ScientificReports_2016}
\begin{barticle}
\bauthor{\bsnm{Zhang}, \binits{C.J.}},
\bauthor{\bsnm{Hua}, \binits{J.F.}},
\bauthor{\bsnm{Xu}, \binits{X.L.}},
\bauthor{\bsnm{Li}, \binits{F.}},
\bauthor{\bsnm{Pai}, \binits{C.-H.}},
\bauthor{\bsnm{Wan}, \binits{Y.}},
\bauthor{\bsnm{Wu}, \binits{Y.P.}},
\bauthor{\bsnm{Gu}, \binits{Y.Q.}},
\bauthor{\bsnm{Mori}, \binits{W.B.}},
\bauthor{\bsnm{Joshi}, \binits{C.}},
\bauthor{\bsnm{Lu}, \binits{W.}}:
\batitle{Capturing relativistic wakefield structures in plasmas using
  ultrashort high-energy electrons as a probe}.
\bjtitle{Scientific Reports}
\bvolume{6},
\bfpage{29485}
(\byear{2016})
\doiurl{10.1038/srep29485}
\end{barticle}
\endbibitem

\bibitem[\protect\citeauthoryear{Wan et~al.}{2024}]{Wan_ScienceAdvances_2024}
\begin{barticle}
\bauthor{\bsnm{Wan}, \binits{Y.}},
\bauthor{\bsnm{Tata}, \binits{S.}},
\bauthor{\bsnm{Seemann}, \binits{O.}},
\bauthor{\bsnm{Levine}, \binits{E.Y.}},
\bauthor{\bsnm{Kroupp}, \binits{E.}},
\bauthor{\bsnm{Malka}, \binits{V.}}:
\batitle{Real-time visualization of the laser-plasma wakefield dynamics}.
\bjtitle{Science Advances}
\bvolume{10}(\bissue{5}),
\bfpage{3595}
(\byear{2024})
\doiurl{10.1126/sciadv.adj3595}
\end{barticle}
\endbibitem

\bibitem[\protect\citeauthoryear{Levine et~al.}{2025}]{Levine_PRR_2025}
\begin{barticle}
\bauthor{\bsnm{Levine}, \binits{E.Y.}},
\bauthor{\bsnm{Wan}, \binits{Y.}},
\bauthor{\bsnm{Tata}, \binits{S.}},
\bauthor{\bsnm{Raspopova}, \binits{D.}},
\bauthor{\bsnm{Kroupp}, \binits{E.}},
\bauthor{\bsnm{Malka}, \binits{V.}}:
\batitle{Direct visualization of shock front induced nonlinear laser wakefield
  dynamics}.
\bjtitle{Phys. Rev. Res.}
\bvolume{7},
\bfpage{012041}
(\byear{2025})
\doiurl{10.1103/PhysRevResearch.7.L012041}
\end{barticle}
\endbibitem

\bibitem[\protect\citeauthoryear{Wan et~al.}{2022}]{Wan_NaturePhysics_2022}
\begin{barticle}
\bauthor{\bsnm{Wan}, \binits{Y.}},
\bauthor{\bsnm{Seemann}, \binits{O.}},
\bauthor{\bsnm{Tata}, \binits{S.}},
\bauthor{\bsnm{Andriyash}, \binits{I.A.}},
\bauthor{\bsnm{Smartsev}, \binits{S.}},
\bauthor{\bsnm{Kroupp}, \binits{E.}},
\bauthor{\bsnm{Malka}, \binits{V.}}:
\batitle{Direct observation of relativistic broken plasma waves}.
\bjtitle{Nature Physics}
\bvolume{18},
\bfpage{1186}--\blpage{1190}
(\byear{2022})
\doiurl{10.1038/s41567-022-01717-6}
\end{barticle}
\endbibitem

\bibitem[\protect\citeauthoryear{Kroupp et~al.}{2020}]{Kroupp_MRE_2022}
\begin{barticle}
\bauthor{\bsnm{Kroupp}, \binits{E.}},
\bauthor{\bsnm{Tata}, \binits{S.}},
\bauthor{\bsnm{Wan}, \binits{Y.}},
\bauthor{\bsnm{Levy}, \binits{D.}},
\bauthor{\bsnm{Smartsev}, \binits{S.}},
\bauthor{\bsnm{Levine}, \binits{E.Y.}},
\bauthor{\bsnm{Seemann}, \binits{O.}},
\bauthor{\bsnm{Adelberg}, \binits{M.}},
\bauthor{\bsnm{Piliposian}, \binits{R.}},
\bauthor{\bsnm{Queller}, \binits{T.}},
\bauthor{\bsnm{Segre}, \binits{E.}},
\bauthor{\bsnm{Phuoc}, \binits{K.T.}},
\bauthor{\bsnm{Kozlova}, \binits{M.}},
\bauthor{\bsnm{Malka}, \binits{V.}}:
\batitle{Commissioning and first results from the new 2 × 100 {TW} laser at
  the {WIS}}.
\bjtitle{Matter and Radiation at Extremes}
\bvolume{7},
\bfpage{044401}
(\byear{2020})
\doiurl{10.1063/5.0090514}
\end{barticle}
\endbibitem

\bibitem[\protect\citeauthoryear{Smartsev et~al.}{2022}]{Smartsev_JoO_2022}
\begin{barticle}
\bauthor{\bsnm{Smartsev}, \binits{S.}},
\bauthor{\bsnm{Tata}, \binits{S.}},
\bauthor{\bsnm{Liberman}, \binits{A.}},
\bauthor{\bsnm{Adelberg}, \binits{M.}},
\bauthor{\bsnm{Mohanty}, \binits{A.}},
\bauthor{\bsnm{Levine}, \binits{E.}},
\bauthor{\bsnm{Seemann}, \binits{O.}},
\bauthor{\bsnm{Wan}, \binits{Y.}},
\bauthor{\bsnm{Kroupp}, \binits{E.}},
\bauthor{\bsnm{Lahaye}, \binits{R.}},
\bauthor{\bsnm{Thaury}, \binits{C.}},
\bauthor{\bsnm{Malka}, \binits{V.}}:
\batitle{Characterization of spatiotemporal couplings with far-field beamlet
  cross-correlation}.
\bjtitle{Journal of Optics}
\bvolume{24},
\bfpage{115503}
(\byear{2022})
\doiurl{10.1088/2040-8986/ac9631}
\end{barticle}
\endbibitem

\bibitem[\protect\citeauthoryear{Andriyash}{2024}]{Andriyash_Axiprop}
\begin{botherref}
\oauthor{\bsnm{Andriyash}, \binits{I.}}:
Axiprop: Simple-to-use Optical Propagation Tool.
\url{https://github.com/hightower8083/axiprop}
\end{botherref}
\endbibitem

\bibitem[\protect\citeauthoryear{Lehe et~al.}{2016}]{Lehe_ComPhysCom_2016}
\begin{barticle}
\bauthor{\bsnm{Lehe}, \binits{R.}},
\bauthor{\bsnm{Kirchen}, \binits{M.}},
\bauthor{\bsnm{Andriyash}, \binits{I.A.}},
\bauthor{\bsnm{Godfrey}, \binits{B.B.}},
\bauthor{\bsnm{Vay}, \binits{J.-L.}}:
\batitle{A spectral, quasi-cylindrical and dispersion-free particle-in-cell
  algorithm}.
\bjtitle{Computer Physics Communications}
\bvolume{203},
\bfpage{66}--\blpage{82}
(\byear{2016})
\doiurl{10.1016/j.cpc.2016.02.007}
\end{barticle}
\endbibitem

\bibitem[\protect\citeauthoryear{Chen}{1983}]{Chen_Plenum_1983}
\begin{bbook}
\bauthor{\bsnm{Chen}, \binits{F.F.}}:
\bbtitle{Introduction to Plasma Physics and Controlled Fusion}.
\bpublisher{Plenum},
\blocation{New York}
(\byear{1983})
\end{bbook}
\endbibitem

\bibitem[\protect\citeauthoryear{Wan et~al.}{2023}]{Wan_2023_LSA_12_116}
\begin{barticle}
\bauthor{\bsnm{Wan}, \binits{Y.}},
\bauthor{\bsnm{Tata}, \binits{S.}},
\bauthor{\bsnm{Seemann}, \binits{O.}},
\bauthor{\bsnm{Levine}, \binits{E.Y.}},
\bauthor{\bsnm{Smartsev}, \binits{S.}},
\bauthor{\bsnm{Kroupp}, \binits{E.}},
\bauthor{\bsnm{Malka}, \binits{V.}}:
\batitle{Femtosecond electron microscopy of relativistic electron bunches}.
\bjtitle{Light: Science \& Applications}
\bvolume{12}(\bissue{1}),
\bfpage{116}
(\byear{2023})
\doiurl{10.1038/s41377-023-01142-1}
\end{barticle}
\endbibitem

\bibitem[\protect\citeauthoryear{Gorbunov and
  Kirsanov}{1987}]{Gorbunov_1987_SPJETP_66_290}
\begin{barticle}
\bauthor{\bsnm{Gorbunov}, \binits{L.M.}},
\bauthor{\bsnm{Kirsanov}, \binits{V.I.}}:
\batitle{Excitation of plasma waves by an electromagnetic wave packet}.
\bjtitle{Soviet Physics JETP}
\bvolume{66},
\bfpage{290}--\blpage{294}
(\byear{1987})
\end{barticle}
\endbibitem

\bibitem[\protect\citeauthoryear{Lu et~al.}{2007}]{Lu_PhysRevSTAB_2007}
\begin{barticle}
\bauthor{\bsnm{Lu}, \binits{W.}},
\bauthor{\bsnm{Tzoufras}, \binits{M.}},
\bauthor{\bsnm{Joshi}, \binits{C.}},
\bauthor{\bsnm{Tsung}, \binits{F.S.}},
\bauthor{\bsnm{Mori}, \binits{W.B.}},
\bauthor{\bsnm{Vieira}, \binits{J.}},
\bauthor{\bsnm{Fonseca}, \binits{R.A.}},
\bauthor{\bsnm{Silva}, \binits{L.O.}}:
\batitle{Generating multi-{GeV} electron bunches using single stage laser
  wakefield acceleration in a {3D} nonlinear regime}.
\bjtitle{Physical Review Special Topics - Accelerators and Beams}
\bvolume{10},
\bfpage{061301}
(\byear{2007})
\doiurl{10.1103/PhysRevSTAB.10.061301}
\end{barticle}
\endbibitem

\bibitem[\protect\citeauthoryear{Smartsev et~al.}{2024}]{Smartsev_OL_2024}
\begin{barticle}
\bauthor{\bsnm{Smartsev}, \binits{S.}},
\bauthor{\bsnm{Liberman}, \binits{A.}},
\bauthor{\bsnm{Andriyash}, \binits{I.A.}},
\bauthor{\bsnm{Cavagna}, \binits{A.}},
\bauthor{\bsnm{Flacco}, \binits{A.}},
\bauthor{\bsnm{Giaccaglia}, \binits{C.}},
\bauthor{\bsnm{Kaur}, \binits{J.}},
\bauthor{\bsnm{Monzac}, \binits{J.}},
\bauthor{\bsnm{Tata}, \binits{S.}},
\bauthor{\bsnm{Vernier}, \binits{A.}},
\bauthor{\bsnm{Malka}, \binits{V.}},
\bauthor{\bsnm{Lopez-Martens}, \binits{R.}},
\bauthor{\bsnm{Faure}, \binits{J.}}:
\batitle{Simple few-shot method for spectrally resolving the wavefront of an
  ultrashort laser pulse}.
\bjtitle{Optics Letters}
\bvolume{49}(\bissue{8}),
\bfpage{1900}--\blpage{1903}
(\byear{2024})
\doiurl{10.1364/ol.502000}
\end{barticle}
\endbibitem

\bibitem[\protect\citeauthoryear{Radier et~al.}{}]{Radier_2022_HPLSE_10}
\begin{botherref}
\oauthor{\bsnm{Radier}, \binits{C.}},
\oauthor{\bsnm{Chalus}, \binits{O.}},
\oauthor{\bsnm{Charbonneau}, \binits{M.}},
\oauthor{\bsnm{Thambirajah}, \binits{S.}},
\oauthor{\bsnm{Deschamps}, \binits{G.}},
\oauthor{\bsnm{David}, \binits{S.}},
\oauthor{\bsnm{Barbe}, \binits{J.}},
\oauthor{\bsnm{Etter}, \binits{E.}},
\oauthor{\bsnm{Matras}, \binits{G.}},
\oauthor{\bsnm{Ricaud}, \binits{S.}},
\oauthor{\bsnm{Leroux}, \binits{V.}},
\oauthor{\bsnm{Richard}, \binits{C.}},
\oauthor{\bsnm{Lureau}, \binits{F.}},
\oauthor{\bsnm{Baleanu}, \binits{A.}},
\oauthor{\bsnm{Banici}, \binits{R.}},
\oauthor{\bsnm{Gradinariu}, \binits{A.}},
\oauthor{\bsnm{Caldararu}, \binits{C.}},
\oauthor{\bsnm{Capiteanu}, \binits{C.}},
\oauthor{\bsnm{Naziru}, \binits{A.}},
\oauthor{\bsnm{Diaconescu}, \binits{B.}},
\oauthor{\bsnm{Iancu}, \binits{V.}},
\oauthor{\bsnm{Dabu}, \binits{R.}},
\oauthor{\bsnm{Ursescu}, \binits{D.}},
\oauthor{\bsnm{Dancus}, \binits{I.}},
\oauthor{\bsnm{Ur}, \binits{C.A.}},
\oauthor{\bsnm{Tanaka}, \binits{K.A.}},
\oauthor{\bsnm{Zamfir}, \binits{N.V.}}:
10 {PW} peak power femtosecond laser pulses at {ELI-NP}
\textbf{10}
\doiurl{10.1017/hpl.2022.11}
\end{botherref}
\endbibitem

\bibitem[\protect\citeauthoryear{Vieira and Mendonça}{2014}]{Vieira_PRL_2014}
\begin{barticle}
\bauthor{\bsnm{Vieira}, \binits{J.}},
\bauthor{\bsnm{Mendonça}, \binits{J.T.}}:
\batitle{Nonlinear laser driven donut wakefields for positron and electron
  acceleration}.
\bjtitle{Physical Review Letters}
\bvolume{112},
\bfpage{215001}
(\byear{2014})
\doiurl{10.1103/PhysRevLett.112.215001}
\end{barticle}
\endbibitem

\bibitem[\protect\citeauthoryear{Mendonça and Vieira}{2014}]{Mendoca_PoP_2014}
\begin{barticle}
\bauthor{\bsnm{Mendonça}, \binits{J.T.}},
\bauthor{\bsnm{Vieira}, \binits{J.}}:
\batitle{Donut wakefields generated by intense laser pulses with orbital
  angular momentum}.
\bjtitle{Physics of Plasmas}
\bvolume{21}(\bissue{3}),
\bfpage{033107}
(\byear{2014})
\doiurl{10.1063/1.4868967}
\end{barticle}
\endbibitem

\bibitem[\protect\citeauthoryear{Thévenet et~al.}{2024}]{thevenet_lasy}
\begin{botherref}
\oauthor{\bsnm{Thévenet}, \binits{M.}},
\oauthor{\bsnm{Andriyash}, \binits{I.A.}},
\oauthor{\bsnm{Fedeli}, \binits{L.}},
\oauthor{\bsnm{Ferran~Pousa}},
\oauthor{\bsnm{Huebl}, \binits{A.}},
\oauthor{\bsnm{Jalas}, \binits{S.}},
\oauthor{\bsnm{Kirchen}, \binits{M.}},
\oauthor{\bsnm{Lehe}, \binits{R.}},
\oauthor{\bsnm{Shalloo}, \binits{R.J.}},
\oauthor{\bsnm{Sinn}, \binits{A.}},
\oauthor{\bsnm{Vay}, \binits{J.-L.}}:
LASY: LAser manipulations made eaSY
(2024).
\url{https://arxiv.org/abs/2403.12191}
\end{botherref}
\endbibitem

\bibitem[\protect\citeauthoryear{Lehe et~al.}{2016}]{Lehe_2016_PRE_94_53305}
\begin{barticle}
\bauthor{\bsnm{Lehe}, \binits{R.}},
\bauthor{\bsnm{Kirchen}, \binits{M.}},
\bauthor{\bsnm{Godfrey}, \binits{B.B.}},
\bauthor{\bsnm{Maier}, \binits{A.R.}},
\bauthor{\bsnm{Vay}, \binits{J.-L.}}:
\batitle{Elimination of numerical cherenkov instability in flowing-plasma
  particle-in-cell simulations by using {Galilean} coordinates}.
\bjtitle{Physical Review E}
\bvolume{94}(\bissue{5}),
\bfpage{053305}
(\byear{2016})
\doiurl{10.1103/physreve.94.053305}
\end{barticle}
\endbibitem

\end{thebibliography}

\begin{appendices}

\section{Extended Data}\label{Extended}

 \begin{figure*}[h]
		\centering
		\includegraphics[width=\linewidth]{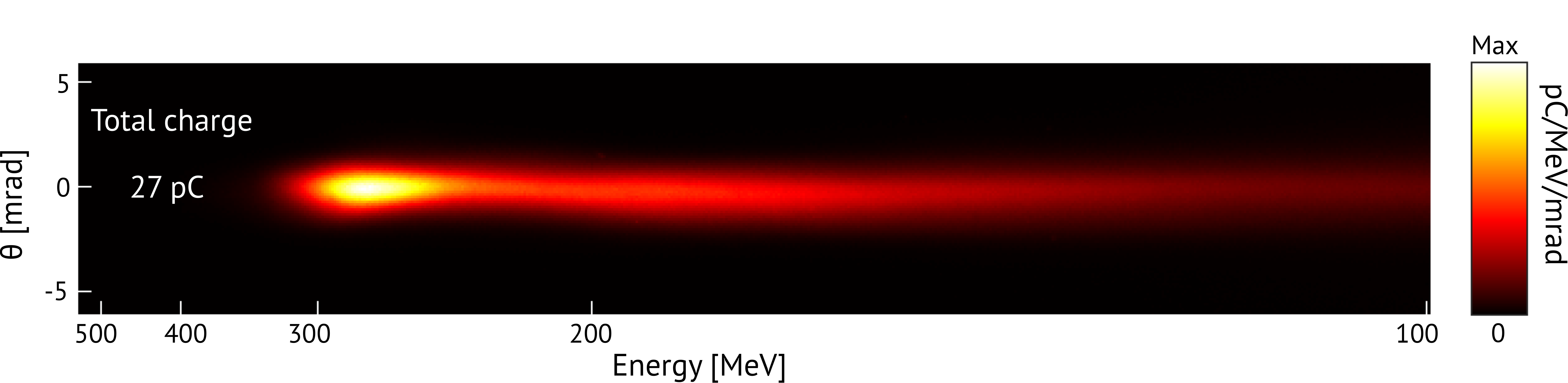}
		\caption{\label{fig:B2_Lanex} \small Sample angularly resolved spectrum of the electrons accelerated by beam 2.}
\end{figure*}

\begin{figure*}[h]
    \centering
    \includegraphics[width=\linewidth]{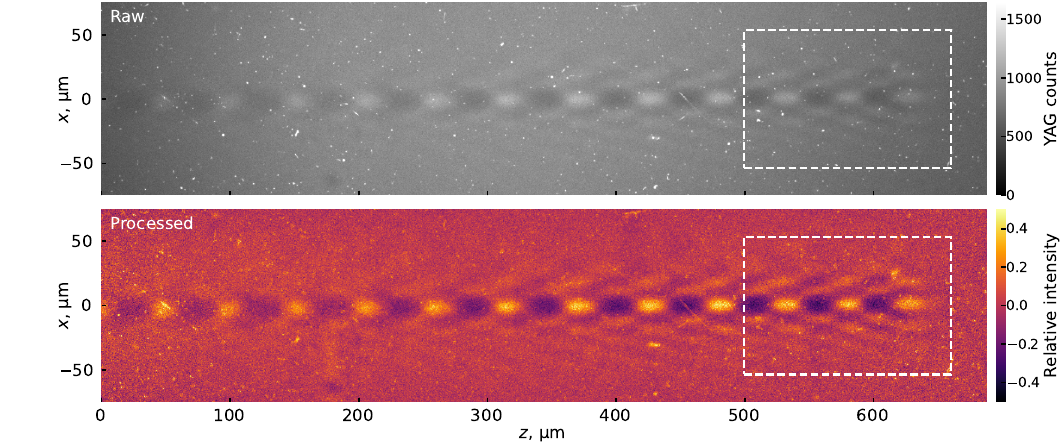}
    \caption{Raw (top) and processed (bottom) FREM images corresponding to the case in figure~\ref{fig:Data}(b).
    The white rectangle shows the crop used in figure~\ref{fig:Data}(b).}
    \label{fig:frem_raw}
\end{figure*}

\begin{figure*}[h]
    \centering
    \includegraphics[]{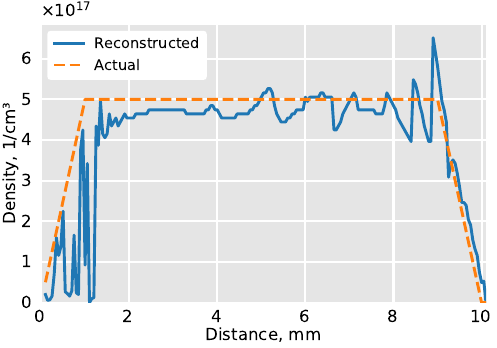}
    \caption{Comparison of density reconstructed from simulated probe images to the density profile used in PIC simulations.}
    \label{fig:density_reconstruction}
\end{figure*}

\begin{figure*}[h]
		\centering
		\includegraphics[width=\linewidth]{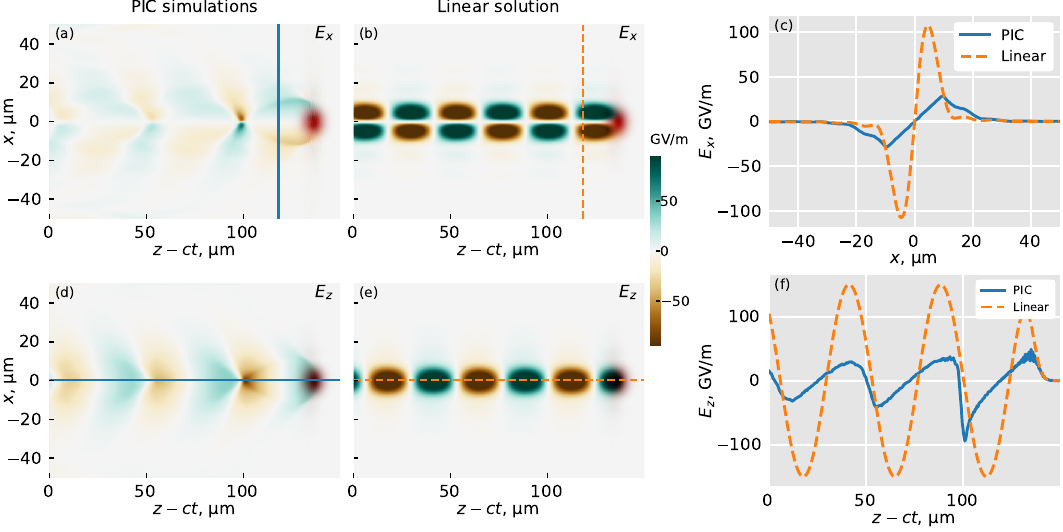}
		\caption{\label{fig:linear_parabola} \small Comparison of the spatial distribution of the transverse electric field $E_x$ between (a) the PIC simulation for the parabola-reflected pulse and (b) the corresponding calculated linear solution. (c) Transverse distribution of $E_x$ for a slice $z - ct = \SI{118}{\um}$ shown with vertical lines in (a--b). Comparison of the longitudinal electric field $E_z$ for (d) the PIC simulation and (e) the linear solution. (f) On-axis longitudinal distributions of $E_z$. The red color in (a--b, d--e) shows the distribution of the laser pulse intensity.}
\end{figure*}

\end{appendices}

\end{document}